\documentclass[11pt]{article}
\RequirePackage{xspace}
\usepackage{a4}
\usepackage{latexsym} 
\usepackage{amssymb}  
\usepackage{amsfonts}  
\usepackage{graphicx}
\usepackage{epsfig}
\usepackage{pslatex}
\usepackage{hyperref}
\usepackage{relsize}
\usepackage{amsmath}
\usepackage[english]{babel}
\usepackage{cite}

\def\auth#1{~\hfill {\it #1}~\\}
\def\spk#1#2{#1\, ({\it #2})}

\def\tit#1{~\\\noindent{\bf #1}}

\newenvironment{wparticipants}{\noindent {\sf PARTICIPANTS:}\\[2mm]}{\\[5mm]}

\newfont{\Thuge}{cmr10 scaled 8000}
\newfont{\thuge}{cmr10 scaled 6000}

\def\mean#1{\ensuremath{\left<#1\right>}}

\newcommand{\gaga}{\gamma\,\gamma}

\providecommand{\ups}{\Upsilon}

\def\ttt#1{\texttt{\small #1}}

\def\beq{\begin{eqnarray}}
\def\eeq{\end{eqnarray}}
\newcommand{\pt}{\ensuremath{p_{T}}\xspace}

\newcounter{zyxabstract}     

\def\pep{PEP-II}
\def\BF{$B$-factory}
\newcommand{\epem}{\mathrm{e}^+\mathrm{e}^-} 
\def\pbar       {\kern 0.2em\overline{\kern -0.2em p}{}\xspace}
\def\lambdac    {\ensuremath{\Lambda_c^+}\xspace}
\def\lambdacbar {\kern 0.2em\overline{\kern -0.2em \Lambda}{}_c^-\xspace}

\def\B       {\ensuremath{B}\xspace}
\def\Bbar    {\kern 0.18em\overline{\kern -0.18em B}{}\xspace}
\def\Bb      {\ensuremath{\Bbar}\xspace}
\def\BB      {\ensuremath{B\Bbar}\xspace} 
\def\piz   {\ensuremath{\pi^0}\xspace}
\def\pip   {\ensuremath{\pi^+}\xspace}
\def\pim   {\ensuremath{\pi^-}\xspace}
\def\pipi  {\ensuremath{\pi^+\pi^-}\xspace}
\def\pipm  {\ensuremath{\pi^\pm}\xspace}
\def\pimp  {\ensuremath{\pi^\mp}\xspace}
\def\kaon  {\ensuremath{K}\xspace}
\def\K     {\ensuremath{K}\xspace}
\def\Kbar  {\kern 0.2em\overline{\kern -0.2em K}{}\xspace}
\def\Kp    {\ensuremath{K^+}\xspace}
\def\Km    {\ensuremath{K^-}\xspace}
\def\Kpm   {\ensuremath{K^\pm}\xspace}
\def\Kmp   {\ensuremath{K^\mp}\xspace}
\def\g     {\ensuremath{\gamma}\xspace}
\def\gaga  {\ensuremath{\gamma\gamma}\xspace}  

\def\qqbar {\ensuremath{q\overline q}\xspace}

\def\bbbar {\ensuremath{b\overline b}\xspace}
\mathchardef\Upsilon="7107
\def\CP                {\ensuremath{C\!P}\xspace}
\def\epem       {\ensuremath{e^+e^-}\xspace}

\begin{document}

\begin{center}
{\Large\bf Parton fragmentation in the vacuum and in the medium} \\
~\\
{\large Mini-proceedings Workshop ECT*, Trento,  Feb. 25 - 29, 2008}\\[0.3cm]

S.~Albino$^{1}$, F.~Anulli$^{2}$, 
F.~Arleo$^{3}$, 
D.~Besson$^{4}$, W.~Brooks$^{5}$, B.~Buschbeck$^{6}$, M.~Cacciari$^{7}$, 
E.~Christova$^{8}$, G.~Corcella$^{9}$, D.~d'Enterria$^{10}$, J.~Dolej\v{s}\'i$^{11}$, 
S.~Domdey$^{12}$, M.~Estienne$^{13}$, K.~Hamacher$^{14}$, M.~Heinz$^{15}$, 
K.~Hicks$^{16}$, D.~Kettler$^{17}$, S.~Kumano$^{18}$, S.-O.~Moch$^{19}$, 
V.~Muccifora$^{20}$, S.~Pacetti$^{20,21}$, R.~P\'erez-Ramos$^{1}$, H.-J.~Pirner$^{12}$, 
A.~Pronko$^{22}$, 
M.~Radici$^{23}$, J.~Rak$^{24}$, C.~Roland$^{25}$, 
G.~Rudolph$^{26}$, Z.~R\'urikov\'a$^{27}$, C.~A.~Salgado$^{28}$, S.~Sapeta$^{29}$,
D.~H.~Saxon$^{30}$, R.~Seidl$^{31}$, R.~Seuster$^{32}$, M.~Stratmann$^{33}$, 
M.~J.~Tannenbaum$^{34}$, M.~Tasevsky$^{11}$, T.~Trainor$^{17}$, D.~Traynor$^{35}$,
M.~Werlen$^{3}$, and C.~Zhou$^{36}$\\[0.5cm]

{\small

{\it $^{1}$Institut f\"ur Theoretische Physik, Universit\"at Hamburg, 
22761 Hamburg, Germany}\\
{\it $^{2}$Sezione di Roma dell'INFN, I-00185 Roma, Italy}\\
{\it $^{3}$LAPTH, Annecy-le-Vieux, France}\\
{\it $^{4}$Kansas University, USA}\\
{\it $^{5}$Dept. F\'{i}sica y Centro de Estudios Subat\'omicos, Univ. T\'ecnica F. Santa Maria, Valpara\'{\i}so, Chile}\\
{\it $^{6}$OA Wien, Austria}\\
{\it $^{7}$LPTHE, UPMC Universit\'e Paris 6, 
CNRS Paris, France}\\
{\it $^{8}$Institute for Nuclear Research and Nuclear Energy, BAS, Sofia, Bulgaria}\\
{\it $^{9}$Museo Storico della Fisica / Centro 
E. Fermi Roma, and Scuola Normale Superiore / INFN, Pisa, Italy}\\
{\it $^{10}$ CERN, CH-1211 Geneva 23, Switzerland}\\
{\it $^{11}$ Charles University, 
Institute of Particle \& Nuclear Physics, 
180 00 Prague 8, Czech Republic}\\
{\it $^{12}$Institut f\"ur Theoretische Physik, Universit\"at Heidelberg, D-69120 Heidelberg, Germany}\\
{\it $^{13}$IPHC, Strasbourg, France}\\
{\it $^{14}$Bergische University Wuppertal, Germany}\\
{\it $^{15}$Yale University, Physics Department, WNSL, 
New Haven, CT 06520, USA}\\
{\it $^{16}$Ohio University, USA}\\
{\it $^{17}$CENPA 354290, University of Washington, Seattle, 98195, USA}\\
{\it $^{18}$Inst. of Particle and Nuclear Studies, KEK, 1-1, Ooho, Tsukuba, Ibaraki, 305-0801, Japan}\\
{\it $^{19}$DESY, Platanenallee 6,  D-15738 Zeuthen, Germany}\\
{\it $^{20}$INFN Laboratori Nazionali di Frascati, via E.Fermi, Frascati, Italy}\\
{\it $^{21}$Centro E.~Fermi, Roma, Italy}\\
{\it $^{22}$Fermilab, P.O. Box 500, MS 318, Batavia, IL, 60510, USA}\\
{\it $^{23}$INFN, Sezione di Pavia, I-27100 Pavia}\\
{\it $^{24}$University of Jyv\"askyl\"a, Finland}\\
{\it $^{25}$MIT, Cambridge, USA}\\
{\it $^{26}$Inst. f. Astro- und Teilchenphysik, Universit\"at Innsbruck, Austria}\\
{\it $^{27}$DESY, Hamburg, Germany}\\
{\it $^{28}$Dept. F\'\i sica de Part\'\i culas and IGFAE, Univ. Santiago de Compostela}\\
{\it $^{29}$M. Smoluchowski Institute of Physics, Jagellonian University, 
30-059 Cracow, Poland}\\
{\it $^{30}$University of Glasgow, Scotland}\\
{\it $^{31}$RIKEN Brookhaven Research Center, Upton, NY 11973, USA}\\
{\it $^{32}$University of Victoria, Canada}\\
{\it $^{33}$Radiation Laboratory, RIKEN, 2-1 Hirosawa, Wako, Saitama, Japan}\\
{\it $^{34}$Brookhaven National Laboratory, Upton, NY 11973, USA}\\
{\it $^{35}$QMWC, Dept. of Physics, London E1 4NS, England}\\
{\it $^{36}$McGill University, Montr\'eal, Canada}\\
}

\vspace{0.4cm}
{\large ABSTRACT}
\end{center}
We present the mini-proceedings of the workshop on ``Parton fragmentation in the vacuum and in the medium'' 
held at the European Centre for Theoretical Studies in Nuclear Physics and Related Areas (ECT*, Trento) 
in February 2008. The workshop gathered both theorists and experimentalists to discuss 
the current status of investigations of quark and gluon fragmentation into hadrons at different accelerator
facilities (LEP, $B$-factories, JLab, HERA, RHIC, and Tevatron) as well as preparations for extension 
of these studies at the LHC. The main physics topics covered were: (i) light-quark and gluon fragmentation 
in the vacuum including theoretical (global fits analyses and MLLA) and experimental (data from 
\epem, $p$-$p$, $e$-$p$ collisions) aspects, (ii) strange and heavy-quark fragmentation, 
(iii) parton fragmentation in cold QCD matter (nuclear DIS), and (iv) medium-modified 
fragmentation in hot and dense QCD matter (high-energy nucleus-nucleus collisions). 
These mini-proceedings consist of an introduction and short summaries of the talks presented at the meeting.\\\\
\vspace{0.4cm}
\noindent\rule{4cm}{0.3pt}\\


\tableofcontents

\vspace{0.4cm}

\section{Introduction}

\noindent
The transition from coloured quarks and gluons to colourless hadrons -- the so-called fragmentation 
or hadronization process -- is a Quantum Chromodynamics (QCD) phenomenon with many important 
theoretical and phenomenological implications for the physics at high-energy colliders. At large hadron 
fractional momenta $z=p_{\rm hadron}/p_{\rm parton}\gtrsim 0.1$, the fragmentation functions (FFs) of 
partons into hadrons or into photons obey DGLAP evolution equations and are obtained from 
global analyses (fits) of various experimental hadron and photon production data. The FFs are, 
for example, a basic ingredient for the calculation of the production of high transverse-momentum 
particles at collider energies within perturbative QCD. At small $z$, successful QCD resummation 
techniques (e.g. the Modified Leading Logarithmic Approximation, MLLA) have been developed to 
understand and describe the evolution of a highly-virtual time-like partons into final hadrons. Most of 
the available data used in the study of the fragmentation process comes from $q\bar{q}$($g$) 
production at $e^+e^-$ colliders (LEP, SLC), although data from deep-inelastic $e$-$p$ scattering (DIS) 
at HERA has also been used especially in the heavy-quark sector. New high-precision flavour-identified 
hadron data from the $B$-factories (BELLE, BABAR), from DIS (HERA), and from hadronic colliders 
(RHIC, Tevatron), can further help to constrain the FFs and, {\em in fine}, to improve the perturbative calculation of 
hadron and photon production at the LHC (in a way similar to the recent developments 
in improved global-analyses of parton distribution functions). In the near future, the vast kinematical range 
opened in $p$-$p$ collisions at LHC energies and luminosities will open up new channels for the study 
of fragmentation functions.\\

\noindent
The study of possible modifications of the parton fragmentation processes in heavy-ion collisions is also 
an important tool for the determination of the thermodynamical and transport properties of the dense 
QCD matter produced at RHIC and LHC energies. Since the discovery of the suppression of high-$p_T$ hadrons 
in central Au-Au collisions at RHIC, a lot of effort has been devoted to the understanding of the propagation 
of partons in QCD media (cold nuclear matter or quark-gluon-plasma) as well as to understand such a 
``medium-modified'' fragmentation mechanism. These phenomenological studies - which often use techniques 
originally developed ``in the vacuum'' (e.g. MLLA) - are well supplemented by a wealth of new data from 
DIS on nuclear targets (HERMES at HERA, and CLAS at JLab) and in heavy-ion collisions (RHIC).\\

A few months before the start of the LHC, it seemed a timely moment to have a workshop, gathering both 
theorists and experimentalists, to discuss the current status of investigations concerning the fragmentation 
mechanisms of gluons, light-quarks and heavy-quarks into hadrons and photons, as well as to discuss their 
implications for the upcoming $p$-$p$ and A-A experimental programmes at the LHC. 
A meeting was organized at the ECT* (Trento) from February 25--29, 2008,  with partial financial support 
from the center. The meeting had 50 participants whose names and institutes are listed in Appendix~A. 
There were 43 presentations of various lengths. Ample time was left for 
discussions after each talk. The talks and discussions were organized around the following main topics:
\begin{itemize}
\item Light-quark and gluon fragmentation in the vacuum (\epem, $p$-$p$, $e$-$p$ collisions): global fits analyses and MLLA.
\item Strange and heavy-quark fragmentation.
\item Parton fragmentation in cold QCD matter. 
\item Medium-modified fragmentation in hot and dense QCD matter.
\end{itemize}

\noindent
This mini-proceedings include a short summary of each talk including 
relevant references, the list of participants and the workshop programme. We felt that such 
a format was more appropriate a format than full-fledged proceedings. Most results are or 
will soon be published and available on arXiv. Most of the talks can also be downloaded 
from the workshop website:\\

\centerline{ \texttt{http://arleo.web.cern.ch/arleo/ff\_vacuum\_medium\_ect08}}

\bigskip

\noindent
We thank the ECT* management and secretariat, in particular Cristina Costa, for the excellent 
organization of the workshop and all participants for their valuable contributions. We believe 
that this was only the first workshop of this kind and look forward to similar meetings in the future.\\

\vspace{0.6cm}

\noindent
{\sc Fran\c{c}ois Arleo, David d'Enterria}

\newpage

\section{Theoretical aspects of parton fragmentation in the vacuum: global fits and MLLA}

\tit{Vacuum fragmentation functions: a common interface?}

\auth{Fran\c{c}ois Arleo, David d'Enterria}


A lot of effort has been invested over the past few years in order to gather all existing sets of 
Parton Distributions Functions (PDFs) under a common interface. This idea was developed in the 
Les Houches Workshop 2001 and later gave birth to the LHAPDF interface~\cite{Giele:2002hx,Whalley:2005nh}, 
aimed to be the successor of PDFLIB~\cite{PlothowBesch:1992qj}. The essential points of LHAPDF, 
summarized in~\cite{Giele:2002hx}, are:
\begin{itemize}
\item PDF sets including their uncertainties should be easily handled, such that any user can 
quantify the influence of PDF uncertainties on a given observable;
\item QCD evolution codes can be compared easily;
\item PDF sets are given under a parametrized form at a soft scale $Q_0$ and not through heavy grids 
(yet this is now possible in practice for older sets);
\item LHAPDF is modular and therefore flexible enough to facilitate the inclusion of new evolution codes or new FF sets.
\end{itemize}

As shown in the workshop, there is also an important ongoing activity in the FF parametrizations community
with e.g. the AKK08, DSS, and HKNS sets recently released, which complement and refine previously
existing analyses (AKK, BFG, BKK, KKP, Kretzer, \dots). All the existing sets with their main characteristics 
are summarized in Table~\ref{tab:ffcharacteristics}.

\begin{table}[htdp]
\begin{center}
\begin{tabular}{p{1.2cm}cp{8.3cm}ccc}
\hline \hline\\
Name & Ref. & Species & Error & $z_{{\rm min}}$ & $Q^2$ (GeV$^2$)\\ \hline \\
AKK & \cite{AKK}  & $\pi^\pm$, $K^\pm$, $K^0_s$, $p$, $\bar{p}$ $\Lambda$, $\bar{\Lambda}$ &  no & 0.1  & 2 -- $4 \cdot 10^4$\\
AKK08 & \cite{AKK08} & $\pi^\pm$, $K^\pm$, $K^0_s$, $p$, $\bar{p}$ $\Lambda$, $\bar{\Lambda}$ &  yes & 0.05  & 2 -- $4 \cdot 10^4$\\
BKK & \cite{BKK} & $\pi^++\pi^-$, $\pi^0$, $K^++K^-$, $K^0+\bar{K^0}$, $h^++h^-$ &  no & 0.05  & 2 -- 200 \\
BFG & \cite{BFGWphot}& $\gamma$ &  no & $10^{-3}$  & 2 -- 1.2 $\cdot 10^4$\\
BFGW & \cite{BFGW} & $h^\pm$ &  yes\footnotemark & $10^{-3}$  & 2 -- 1.2 $\cdot 10^4$\\
CGRW & \cite{Chiappetta:1992uh} & $\pi^0$ &  no & $10^{-3}$  & 2 -- 1.2 $\cdot 10^4$\\
DSS & \cite{DSS1,DSS2} & $\pi^\pm$, $K^\pm$, $p$, $\bar{p}$, $h^\pm$ &  yes\footnotemark & 0.05-0.1  & 1 -- $10^5$ \\ 
DSV & \cite{DSV} & polarized and unpolarized $\Lambda$ &  no & 0.05  & 1 -- $10^4$ \\ 
GRV & \cite{Gluck:1992zx} & $\gamma$ &  no & 0.05  & $\ge$ 1 \\ 
HKNS & \cite{HKNS} & $\pi^\pm$, $\pi^0$, $K^\pm$, $K^0+\bar{K^0}$, $n$, $p+\bar{p}$ &  yes & $0.01$ -- $1$ & $1$ -- $10^8$ \\
KKP & \cite{KKP} & $\pi^++\pi^-$, $\pi^0$, $K^++K^-$, $K^0+\bar{K^0}$, $p+\bar{p}$, $n+\bar{n}$, $h^++h^-$  &  no & 0.1  & 1 -- $10^4$ \\
Kretzer & \cite{Kretzer} & $\pi^\pm$, $K^\pm$, $h^++h^-$ &  no & 0.01  & 0.8 -- $10^6$ \\&&&&&\\
\hline \hline
\end{tabular}
\end{center}
\label{tab:ffcharacteristics}
\caption{Main characteristics of fragmentation function (FF) sets obtained from global fit analyses.}
\end{table}%

\footnotetext{Errors on the parameters and their correlation are given.}
\footnotetext{Typical uncertainties for ``truncated'' energy fractions $\int_{0.2}^1 z D_i^h(z,Q^2) dz$ studied for all flavors and 
hadron species using the Lagrange multiplier method.}

Just like parton densities, the first error analyses in the fragmentation functions have also just appeared. 
In this context, it appears sensible to examine whether 
an effort similar to LHAPDF should be performed for the different FF sets. Ideally, a common interface 
would facilitate the systematic comparison between the different sets, in order to explore for instance 
the theoretical uncertainties of NLO calculations for hadron production in $p$-$p$ collisions at the LHC associated 
with the FFs (which are usually much larger than the uncertainties due to the PDFs). As a very first step, 
a simple wrapper program should be developed to call any parton-to-hadron FF for any given species. 
Also, the latest produced FF parametrizations should be made available in a common public web-page such 
as the existing one from Marco Radici~\cite{Radiciweb}.


\tit{Albino-Kniehl-Kramer FFs: Improvements from new theoretical input and experimental data}

\auth{Simon~Albino}

We present results from an update (AKK08)~\cite{AKK08} 
of the previous AKK fragmentation function (FF) sets~\cite{AKK} for charged
pions, charged kaons, (anti)protons, neutral kaons and (anti)lambdas
with additional theoretical and experimental input. 
We incorporate hadron mass effects~\cite{Albino:2005gd}, and fit the
hadron mass in the case of the \epem calculation which also has the
effects of subtracting out other low $\sqrt{s}$ and small-$x$ effects
beyond the fixed order approach, such as higher twist, small-$x$
logarithms, etc. In the case of the baryons, the fitted masses are
about 1\% above the true masses, which is consistent with scenarios in
which the baryons are produced mainly from direct partonic
fragmentation with a small contribution from decays from slightly
heavier resonances.  A greater excess is found for the pion mass,
suggesting contributions to the sample from decays of heavier
particles such as $\rho(770)$. Charged and neutral kaon masses are
significantly below their true masses.  A possible explanation for
this is that there are significant contributions from complicated
decay channels, such that the direct partonic fragmentation approach
is insufficient. For this reason we do not impose SU(2) isospin
symmetry of $u$ and $d$ quarks between charged and neutral kaons,
which we do for pions.  We implement large-$x$ resummation in the
quark coefficient function of $e^+ e^-$ reactions using the results
from Ref.~\cite{Cacciari:2001cw}, since this is a simple improvement
which modifies the cross section over the whole range in $x$ that we
constrain.  Large-$x$ resummation is also implemented in the DGLAP
evolution of the FFs~\cite{AKK08}.  This results in a
significant improvement in the fit for charged kaons, (anti)protons
and (anti)lambdas, while $\chi^2$ is essentially unchanged for neutral
kaons and charged pions.\\

In addition to the constraints of the previous AKK fit, we impose
further constraints on the charge-sign unidentified FFs from the data
for single inclusive production of identified particles ($p_T \geq 2$GeV/c) 
from RHIC~\cite{Arsene:2007jd,Adler:2003pb,Adams:2006nd,Adams:2006uz,Abelev:2006cs},
the Tevatron~\cite{Acosta:2005pk} and
electron-positron reactions below the Z pole mass and in the range
$0.05<x<0.1$.  While the untagged measurements from electron-positron
reactions provide excellent constraints for the sums of the
charge-sign unidentified FFs for quarks of the same electroweak
charges, they do not constrain the remaining degrees of freedom at
all. As in the previous AKK fit, these were constrained using quark
tagged data, while in the new AKK fit additional constraints are
provided by the data from RHIC.  These data are also much more
sensitive to gluon fragmentation and impose (exclusively) new
constraints on the charge-sign asymmetry FFs.  Normalization errors
were treated as systematic effects, i.e.\ these errors were
incorporated via a correlation matrix. Their weights were fitted
analytically and independently of the fit in order to further
ascertain the quality of the fit, and their magnitudes 
were typically found to lie in the reasonable range of 0 -- 2.\\

While the results for the fitted masses suggest that the baryons are
the best candidates for studying direct partonic fragmentation, there
unfortunately exist some inconsistencies between the calculation and
the measurements of the inclusive production of these particles at
RHIC: The description of the STAR data for
$\Lambda/\overline{\Lambda}$ fails, while the contribution from the
initial protons' valence $d$ quarks to the charge-sign asymmetry for
$p/\overline{p}$ from STAR is negative. Furthermore, while the
contributions to the production from the fragmentation of the valence
and sea quarks of the initial protons for the data from RHIC exhibited
the expected behaviour, the contribution from the valence $d$ quark
fragmentation to the charge-sign asymmetry in (anti)proton production
is negative. All these issues would be better understood in the
context of an error analysis of the FFs.\\

We compared our sets to the recent DSS~\cite{DSS1} / DSV~\cite{DSV} and 
HKNS~\cite{HKNS} ones, and typically found reasonable agreement for favoured
 FFs but not for unfavoured, and large discrepancies exist at large-$x$ in some cases.
Future hadron identified data from BABAR, CLEO, HERA and RHIC will help
to clarify these issues and significantly reduce the large
uncertainties in much of the FF degrees of freedom.

\tit{De Florian-Sassot-Stratmann (DSS) global QCD analysis of fragmentation functions}

\auth{Marco Stratmann}

We present new sets fragmentation functions for pions, kaons~\cite{DSS1},
protons, and unidentified charged hadrons~\cite{DSS2} obtained in NLO combined
analyses of single-inclusive hadron production in electron-positron annihilation,
proton-proton collisions, and deep-inelastic lepton-proton scattering. 
At variance with previous fits, the present analyses take into
account data where hadrons of different electrical charge are identified,
which allow to discriminate quark from anti-quark fragmentation functions
without the need of non-trivial flavor symmetry assumptions.
The resulting sets are in good agreement with all data analyzed.
The success of the global analysis performed here,
including observables other than $e^+e^-$ annihilation, stands
for an explicit check of factorization, universality, and the underlying
framework of perturbative QCD.\\

Increasingly accessible hadron production data from proton-proton collisions
and hadron multiplicities coming from semi-inclusive DIS should not be disregarded as they
offer a crucial piece of complementary information that reduces
significantly the uncertainties of the resulting fragmentation functions.
For instance, we find that the new DIS pion and kaon multiplicities provided
by the HERMES experiment effectively constrain the separation between favored
and unfavored distributions, a separation that was either not implemented in previous
sets or it was based on certain assumptions.
The most recent RHIC results provide stringent constraints on the
gluon fragmentation function and, in general, on the large $z$ behavior
of the other distributions.\\

The implementation of the $\chi^2$ minimization in our global analysis is
numerically fast and efficient and can be straightforwardly expanded to any
future set of hadron production data. With the help of the Mellin moment technique
~\cite{ref:mellin}, the entire analysis was consistently performed at NLO accuracy without
resorting to often used approximations for NLO hard scattering cross sections.
For completeness we also provide LO sets.
An extensive use of the Lagrange multiplier technique~\cite{ref:lagrange} is made in order to
assess the typical uncertainties in the extraction of the fragmentation functions and
the synergy from the complementary data sets in our global analysis.

\tit{Hirai-Kumano-Nagai-Sudoh (HKNS) fragmentation functions and proposal for exotic-hadron search}

\auth{Shunzo Kumano}

Fragmentation functions and their uncertainties are determined for pions, kaons,
and protons by a global $\chi^2$ analysis of charged-hadron production data in electron-positron 
annihilation and by the Hessian method for error estimation~\cite{HKNS}.
The results indicate that the fragmentation functions, especially gluon and
light-quark fragmentation functions, have large uncertainties
at small $Q^2$. There are large differences between widely-used functions
by KKP (Kniehl, Kramer, and P\"otter)~\cite{KKP}, AKK (Albino, Kniehl, and Kramer)~\cite{AKK},
and Kretzer~\cite{Kretzer}; however, they are compatible with each other and also
with our functions if the uncertainties are taken into account.
We find that determination of the fragmentation functions is improved in
next-to-leading-order (NLO) analyses for the pion and kaon in comparison
with leading-order ones. Such a NLO improvement is not obvious
in the proton. Since the uncertainties are large at small $Q^2$,
the uncertainty estimation is very important for analyzing hadron-production
data at small $Q^2$ or $p_T$ ($Q^2, \, p_T^2 \gg M_Z^2$) in lepton scattering
and hadron-hadron collisions. A code is available for general users
for calculating the obtained fragmentation functions~\cite{HKNS}.\\

Next, it is proposed that fragmentation functions should be used to identify
exotic hadrons~\cite{kumano2}. As an example, fragmentation functions of the scalar meson
$f_0(980)$ are investigated. The $f_0$ meson is considered as a candidate
for an exotic hadron beyond the usual $q\bar q$ configuration because
its strong-decay width is much larger than the experimental one~\cite{kumano3}
if it is an ordinary $q\bar q$ meson.
The radiative decay $\phi \rightarrow f_0 \gamma$ was
proposed to find its internal structure~\cite{kumano4}, and subsequent measurements
indicated tetra-quark or $K\bar K$ structure. However, its structure has not
been clearly determined yet. Here, we investigate the possibility of
ordinary $q\bar q$, $s\bar s$, tetra-quark, $K\bar K$ molecule,
or glueball via the fragmentation functions of $f_0$.
It is pointed out that the second moments
and functional forms of the $u$- and $s$-quark fragmentation functions
can distinguish the tetra-quark structure from $q\bar q$. By the global
analysis of $f_0 (980)$ production data in electron-positron annihilation,
its fragmentation functions and their uncertainties are determined.
It is found that the current available data are not sufficient to
determine its internal structure, while precise data in future should
be able to identify exotic quark configurations.

\tit{Main sources of uncertainty in quark and gluon fragmentation functions into hadrons and photons}

\auth{Monique Werlen}

A precise knowledge of the QCD hadron and photon production is necessary while searching
for e.g. a low mass Higgs signal at LHC and Tevatron. Uncertainties on this background must 
therefore be asserted with care, in particular those due to the measurement of fragmentation 
functions. In the calculation of inclusive hadron and $\gamma$ cross sections in $pp$ collisions
with the PHOX package -- a set of NLO parton-level event generators for large $p_T$ PHOton, 
hadron and/or jet X-sections \cite{phoxfamily} -- uncertainties in fragmentation functions (FFs) 
do matter compared to those from scales and parton distribution functions (PDF) at RHIC
and LHC energies \cite{Arleo:2004gn}. FFs are obtained from fits to inclusive cross sections 
assuming a functional form at a given scale $M_{F0}$ and evolving it to the scale of the data.
The sources of uncertainties are both experimental and theoretical.
Experimental uncertainties are due to statistical errors, in particular at large momentum 
fraction $z$ of the hadron in $e^+e^-$ data. But also systematic effects contribute such as 
in the data normalization or in the extrapolation (evolution) needed to cover the full $z$ ($Q^2$) range.
Matching theory and data due to binning and cuts settings may also be an issue.
Theoretical uncertainties come from the choice of the functional form for the selected $z$ range, 
the choice of scales, the order of the theory (leading, next-to-leading NLO, resummation) 
and further parameters like PDFs and $\alpha_s$.\\

As shown for the first time in \cite{Chiappetta:1992uh}, NLO fits to $e^+ e^- \to \pi^0 X$ 
data from PEP and PETRA constrain the quark fragmentation into neutral pions while 
those to $p p \to \pi^0 X$ from ISR and UA2 constrain the gluon fragmentation.
BFGW \cite{BFGW} sets of FF into unidentified charged hadrons have been obtained from LEP 
and PETRA data
choosing optimized scales: a rough approximation of the large-$z$ resummation (\cite{Cacciari:2001cw}).
Gluon parameters are quite sensitive to the functional form chosen for the quark distribution.
A full statistical error analysis is performed 
while estimation of the ``theoretical error'' as previously detailed can be obtained by comparison 
to other FF sets. Important discrepancies with the BKK set~\cite{BKK} have been shown, specially 
at large $z$ 
while the gluon FFs at $z>0.5$ is found much higher, in accordance with the UA1 data, 
than in the Kretzer set \cite{Kretzer}. Large scale instabilities (specially at low $\sqrt{s}$ and low $p_T$)
affect the phenomenology of inclusive production of pion \cite{Aurenche:1999nz}.
Theoretical estimates rely on extrapolations of the FFs outside ($0.75<x<0.9$) of the region 
where they are actually constrained ($0.1<z<0.7$) by the data~\cite{Binoth:2002ym}. 
A preliminary study~\cite{inprep} with JETPHOX shows that the $p_T$ imbalance hadron-jet 
correlation cross-sections at RHIC energy may be used to constrain the FFs into hadrons in
the high $z$ region. For $p_T(h) > 25$~GeV/c and $p_T (jet) > 30$~GeV/c, the three set 
(BFGW, KKP, Kretzer) give quite different predictions: BFGW is systematically
higher than KKP while Kretzer is a factor two lower.\\


Thanks to recent data from PHENIX and D0 spanning a large $x_T=2p_T/\sqrt{s}$ range,
inclusive prompt photons cross sections from $ \sqrt{s}=23$ GeV to $\sqrt{s}= 1.96$ TeV
are now well understood in the NLO QCD framework \cite{Aurenche:2006vj}.
Photons can be produced either directly or via a parton-to-photon FF. The latter
becomes important at low $p_T$ and high $\sqrt{s}$. Uncertainties on the parton-to-photon
FFs are mainly due to the uncertainty on the gluon FF as quantified by the BFG sets I 
and II~\cite{BFGWphot}. BFG-I reduces the cross section by up to 10\% (a factor 2.5) 
at $p_T$~=~3~GeV at RHIC (LHC) energy compared to BFG II \cite{Arleo:2004gn}.
As shown in \cite{inprep} the photon-jet correlation at RHIC energy may also be used to 
constrain the photon FF. Fixing the jet and varying the photon momenta allows for a direct 
measurement of the photon fragmentation at low $z$, a region barely accessible in the LEP 
experiments.\\

In the calculation of inclusive hadron and photon cross sections in $pp$ collisions
with the PHOX programs -- NLO event generators (parton level) for large $p_T$ 
PHOton (hadron or jet) cross-sections (X-sections)~\cite{phoxfamily} --, 
uncertainties on fragmentation functions (FF) do matter compared
to those from scales and parton distribution functions (PDF) at RHIC
and LHC energies~\cite{Arleo:2004gn}.
There are various sources of uncertainties in the determination of FF from fits to inclusive
cross sections where one assumes a functional form at a scale $M_{F0}$
and evolves it to the scale of the data. Experimental uncertainties are due to statistical errors
(FF with a large momentum fraction $z$ of the parton in $e^+e^-$ data),
systematic, normalization, the probed
z and $Q^2$ range as well as the sensitivity to quark and gluon FF.
Matching theory and data (binning, cuts) may also be an issue.
Theoretical uncertainties come from the choice of the functional form,
the $z$ range of assumptions, the choice of scales, the order of the theory
(leading, next-to-leading NLO, resummation) and further parameters
like PDFs and $\alpha_s$.\\

NLO fits  to $e^+ e^- \to \pi^0 X$ data from PEP and PETRA
constrain the quark fragmentation into neutral pions while NLO fits to $p p \to \pi^0 X$
from ISR and UA2 constrain the gluon fragmentation~\cite{Chiappetta:1992uh}.
BFGW~\cite{BFGW} sets of FF into
unidentified charged hadrons have been obtained from LEP and PETRA data
choosing optimized scales
(a rough approximation of the large $z$ resummation performed later in~\cite{Cacciari:2001cw}).
Gluon parameters were quite sensitive to the functional form
chosen for the quark distribution.
A full statistical error analysis is performed
while estimation of the ``theoretical error'' embedded in the parametrizations
and  which come from various theoretical assumptions
can be obtained by comparing with different sets:
There are important discrepancies, specially at large z, with the BKK set~\cite{BKK} 
while the very low gluon
FF at $z>0.5$ from the Kretzer set~\cite{Kretzer} seems unfavored by
UA1 data. Large scale instabilities (specially at low $\sqrt{s}$ and low $p_T$)
affect the phenomenology of inclusive production of pion~\cite{Aurenche:1999nz}.
Theoretical estimates rely on extrapolations
of the FF outside ($0.75-0.9$) of the region where they
are actually constrained  by the data ($0.1<z<0.7$) from which they are extracted
~\cite{Binoth:2002ym}.
A preliminary study~\cite{inprep} with JETPHOX shows that hadron-jet correlation
at RHIC energy may be used to constrain the FF into hadron in
the high-$z$ region. With $p_T(h) > 25$~GeV/c and $p_T (jet) >30$~GeV/c,
BFGW is systematically higher than KKP~\cite{KKP} while Kretzer is almost a factor two lower
than BFGW.\\

Thanks to recent data from PHENIX and D0 spanning a large $x_T=2p_T/\sqrt{s}$ range,
inclusive prompt photons cross sections from $ \sqrt{s}=23$~GeV to $\sqrt{s}$~=~1.96~TeV
are now well understood in the NLO QCD framework~\cite{Aurenche:2006vj}.
Photons can be produced either directly or via a parton-to-photon FF. The later
become important at low $p_T$ and high $\sqrt{s}$. Uncertainties on the photon
FF are mainly due to the uncertainty on the gluon FF
as quantified by the BFG sets I and II~\cite{BFGWphot}. BFG I
reduce the cross section by up to 10\% (a factor 2.5) at $p_T$~=~3~GeV/c
at RHIC (LHC) energy compared to BFG II~\cite{Arleo:2004gn}.
A preliminary study~\cite{inprep} with JETPHOX  shows that photon-jet correlation
at RHIC energy may be used to constrain the photon FF.
With the momentum of the jet fixed, varying the momentum of the non
isolated photon allows for a direct access to the photon fragmentation at low z,
a region barely accessible in the LEP experiments.


\tit{Time-like splitting functions at NNLO in QCD}

\auth{Sven-Olaf Moch}

We have employed relations between space-like and time-like deep-inelastic processes
in perturbative QCD to calculate the next-to-next-to-leading order (NNLO)
contributions to the time-like non-singlet and singlet quark-quark and gluon-gluon
splitting functions for the scale dependence (evolution) of the
parton fragmentation distributions $\,D_{\! f}^{\,h}(x,Q^2)$.
The evolution equations read
\begin{equation}
\label{eq:Devol}
  {d \over d \ln Q^2} \; D_{i}^{\,h} (x,Q^2) \:\: = \:\:
  \int_x^1 {dz \over z} \; P^{\,T}_{ji} \left( z,\alpha_{\rm s} (Q^2) \right)
  \:  D_{j}^{\,h} \Big( {x \over z},\, Q^2 \Big)\, ,
\end{equation}
where $x$ denotes the fraction of the momentum of the final-state parton $f$ carried by
the outgoing hadron $h$, $\,Q^2$ is a time-like hard scale and
summation over $\,j \, =\, q,\:\bar{q},\:g\,$ is understood.
The time-like splitting functions $\,P^{\,T}_{ji}\,$ governing Eq.~(\ref{eq:Devol})
admit an expansion in powers of the strong coupling $\alpha_{\rm s}$,
\begin{equation}
\label{eq:PTexp}
  P^{\,T}_{ji} \left( x,\alpha_{\rm s} (Q^2) \right) \:\: = \:\:
  a_{\rm s} \, P_{ji}^{(0)\,T}(x) \: + \: a_{\rm s}^{\:\!2} \, P_{ji}^{(1)\,T}(x)
  \: +\: a_{\rm s}^{\:\!3} P_{ji}^{(2)\,T}(x) \: +\: \ldots \:\: ,
\end{equation}
where we normalize to $\,a_{\rm s} \equiv \alpha_{\rm s}(Q^2)/ (4\pi)$.
The leading-order (LO) terms in Eq.~(\ref{eq:PTexp}) are identical to the
space-like case of the initial-state parton distributions,
which is the so-called Gribov-Lipatov relation~\cite{Gribov:1972ri}.
Also at the next-to-leading order (NLO), the functions $\,P_{ji}^{(1)\,T}\!(x)$
are related to their space-like counterparts by a suitable analytic
continuation, see e.g.~\cite{Stratmann:1996hn}.
Following these ideas, we have derived at NNLO the
three distinct non-singlet functions $P^{\,(2),T}_{\rm ns, \xi}$ (with $\xi = \pm,\rm v$)~\cite{Mitov:2006ic}
and the diagonal singlet quantities $P_{\rm qq}^{\,(2),T}$, $P_{\rm gg}^{\,(2),T}$~\cite{Moch:2007tx}
from the corresponding space-like results~\cite{Moch:2004pa} and~\cite{Vogt:2004mw}
using two independent methods.
One approach relied on mass factorization and the structure of the infrared singularities, the other
implemented the concept of universal (kinematics independent) splitting functions as conjectured in~\cite{Dokshitzer:2005bf}.
The off-diagonal three-loop time-like quantities $P_{\rm gq}^{\,(2),T}$, $P_{\rm qg}^{\,(2),T}$ are
presently still unknown.

\tit{Theory of (extended) dihadron fragmentation functions}

\auth{Marco Radici}

Dihadron Fragmentation Functions (DiFF) describe the probability that a quark
hadronizes into two hadrons plus anything else, i.e.\ the process
$q\to h_A\,h_B\,X$. They can appear in lepton-lepton, lepton-hadron and
hadron-hadron collisions semi-inclusively producing final-state hadrons.
A thorough investigation of their formal properties was performed
in the last decade, identifying the leading~\cite{pv1} and
sub-leading twist~\cite{ioale-tw3} terms, and recognizing the possibility of
separating the diagonal and interfering contributions for the $(h_A, h_B)$ pair being
in different relative partial waves~\cite{ioale-LM}.\\

At present, the most important application deals with DiFF as analyzers of the
fragmenting quark spin and regards the extraction of the transversity distribution
$h_1$ in the nucleon~\cite{pv2}, whose knowledge is a basic test of QCD in the
nonperturbative domain~\cite{barone}. The extraction proceeds via the measurement of
a single-spin asymmetry, that has been recently performed by the HERMES~\cite{hermes}
and COMPASS~\cite{compass} collaborations. Information on DiFF can be extracted by
the $e^+e^- \to (h_{A1} h_{A2})\,(h_{B1} h_{B2})\,X$ process~\cite{pv3}, which is
currently being measured by the BELLE collaboration~\cite{belle}, or by
models~\cite{ioale-model}.\\

Recently, DiFF have been recognized as a crucial ingredient for canceling all
collinear singularities in the NLO calculation of the cross section for
$e^+ e^- \to h_A\,h_B\,X$~\cite{deflo}. When DiFF depend only on the
fractional energies of the two final hadrons, there is no way to distinguish
between the usual $q\to h_A\,h_B\,X$ process and the mechanism
$q\to q_A\,q_B \to h_A\,h_B\,X$, where the initial parton branches into two other
partons $q_A,\,q_B,$ each one subsequently hadronizing in one single hadron. The
net outcome is that evolution equations for DiFF contain an inhomogeneous term
involving single-hadron fragmentation functions~\cite{deflo}.\\

However, most of the experimental information about DiFF consists of spectra in the
invariant mass $M_h^2$ of the hadron pair. We keep the explicit dependence
upon $M_h^2$ inside DiFF and we define the so called extended DiFF (extDiFF). Using
the jet calculus technique, we show that this (soft) scale breaks the degeneracy
between $q\to h_A\,h_B\,X$ and $q\to q_A\,q_B \to h_A\,h_B\,X$ mechanisms, producing
for the extDiFF the usual DGLAP evolution~\cite{ioale-evo}. Therefore, we deduce that
the NLO cross section for the $e^+ e^- \to h_A\,h_B\,X$ process, when differential
also in $M_h^2$ and for the hadron pair belonging to the same jet, can be described 
as the factorized convolution of extDiFF and of the
same kernel used for the single-hadron fragmentation case. Finally, we show that it
is possible to separately study the evolution of each single term in the expansion
in partial waves, and we give a preliminary numerical example.


\tit{Fragmentation functions in $e^+e^-$  collisions at low $Q^2$}

\auth{David Kettler}

We report measurements of transverse momentum $p_T$ spectra for ten event multiplicity 
classes of $p$-$p$ collisions at $\sqrt{s} = 200$~GeV at STAR. By analyzing the multiplicity 
dependence we find that the spectrum shape can be decomposed into a part with amplitude 
proportional to multiplicity and described by a L\'evy distribution on transverse mass $m_T$, 
and a part with amplitude proportional to multiplicity squared and described by a Gaussian 
distribution on transverse rapidity $y_T$. The functional forms of the two parts are nearly 
independent of event multiplicity. The two parts can be identified with the soft and hard 
components of a two-component model of $p$-$p$ collisions. This analysis then provides the 
first isolation of the hard component of the $p_T$ spectrum as a distribution of simple form on $y_T$. \\

We analyze the energy scale dependence of fragmentation functions from $e^+e^-$ 
collisions using conventional momentum measures $x_p$ and $\xi_p$ and rapidity $y$. 
We find that replotting fragmentation functions on a normalized rapidity variable results in 
a compact form precisely represented by the beta distribution, its two parameters varying 
slowly and simply with parton energy scale $Q$.  Dijet multiplicities are used to constrain 
these parameters at lower energies.  The resulting parameterization enables extrapolation 
of fragmentation functions to low $Q$ in order to describe fragment distributions at low 
transverse momentum $p_T$.  These results compared favorably to conventional representations of fragmentation functions. \\

Finally, we analyze the minimum bias two-particle correlations of $p$-$p$ collisions.   Cuts are 
used to isolate the jet-like peak in the angular correlations and the corresponding two-particle 
$y_T$ correlations are compared to the fragmentation function parametrization derived from 
$e^+e^-$ collisions.

\newpage

\tit{Jet fragmentation at MLLA and beyond}

\auth{Redamy P\'erez-Ramos}

The hadronic $k_\perp$-spectrum inside a high energy jet is determined including corrections of relative magnitude
 ${\cal O}(\sqrt{\alpha_s})$ with respect to the Modified Leading Logarithmic Approximation (MLLA), 
in the limiting spectrum approximation (assuming an infrared cut-off $Q_0 =\Lambda_{QCD}$) and beyond 
($Q_0\ne\Lambda_{QCD}$). The results in the limiting spectrum approximation are found to be in
impressive agreement with preliminary measurements by the CDF collaboration, unlike
what occurs at MLLA, pointing out small overall non-perturbative contributions.
Within the same framework, 2-particle  correlations inside a jet
are also predicted at Next-to-MLLA and compared to previous MLLA calculations.
MLLA corrections, of relative magnitude ${\cal O}(\sqrt{\alpha_s})$ with respect
to the leading double logarithmic approximation (DLA), were shown to be quite
substantial for single-inclusive distributions and 2-particle correlations
~\cite{PerezMachet,RPR2}. Therefore, it appears legitimate
to wonder whether corrections of order ${\cal O}(\sqrt{\alpha_s})$, that is
next-to-next-to-leading or next-to-MLLA (NMLLA), are negligible or not.\\

The starting point of this analysis is the MLLA evolution equation for
the generating functional of QCD jets~\cite{Basics}.
Together with the initial condition at threshold, it determines
jet properties at every energies. At high energies one can represent
the solution as an expansion in $\sqrt{\alpha_s}$. Then,
the leading (DLA) and next-to-leading (MLLA) approximations are complete.
The next terms (NMLLA) are not complete but they include an important
contribution which takes into account energy conservation and an
improved behavior near threshold.
Some results for such NMLLA terms have been
studied previously for global observables and have been found to
better account for recoil effects. They were shown to
drastically affect multiplicities and particle correlations in jets:
this is in particular the case in ~\cite{CuypersTesima}, which 
deals with multiplicity correlators of order 2, and in~\cite{DokKNO},
where  multiplicity correlators involving a higher number of partons are
studied; in particular, the higher this number, the larger turn out to be  NMLLA corrections.\\

The present study makes use of this evolution equation to estimate NMLLA
contributions to our differential observables.
It presents the complete calculations of the single
inclusive $k_\perp$ distribution leading to the results
published in~\cite{RedamyNMLLA}, and extends them to 2-particle correlations
inside a high energy jet.
The obtained agreement between the CDF results
and the NMLLA distributions over
the whole $k_\perp$-range is excellent.
The NMLLA calculation is in particular able to capture the shape
of CDF spectra at every jet hardness $Q$. Conversely, 
predictions at MLLA prove
only reliable at not too large $k_\perp$.
The domain of validity of the predictions has been enlarged
to larger $k_\perp$ (and thus to larger $x$ since $Y$ is fixed)
computing from MLLA to NMLLA accuracy. This agreement further supports
the Local Hadron Parton Duality (LPHD) hypotheses~\cite{LPHD}.\\

The MLLA, NMLLA equations for correlators were analyzed and solved
iteratively~\cite{RPR2}. This allowed us to generalize the result previously
obtained by Fong and Webber in~\cite{FW} that was valid in the vicinity
of the maximum of the single inclusive parton energy distribution
(``hump'').
In particular, we have analyzed the regions of moderately small $x$
above which the correlation becomes ``negative'' (${\cal C}-1<0$).
This happens when suppression because of the limitation of the
phase space takes over the positive correlation due to gluon cascading.
This region turned out to be larger in NMLLA than in MLLA.
Also, the correlation vanishes (${\cal C}\to1$) when one of the partons
becomes very soft ($\ell = \ln1/x \to Y=\ln E\Theta/Q_0$). The reason
for that is dynamical rather than kinematical: radiation of a soft
gluon occurs at {\em large angles}\/ which makes the radiation
coherent and thus insensitive to the internal parton structure of the
jet ensemble.
Qualitatively, our MLLA and NMLLA result agrees better with available OPAL data
than the Fong--Webber prediction. There remains however a significant
discrepancy, markedly at very small $x$. In this region
non-perturbative effects are likely to be more pronounced. They may
undermine the applicability {\em to particle correlations}\/ of the local
parton--hadron duality considerations that were successful in
translating parton level predictions to hadronic observations in the
case of more inclusive {\em single particle energy spectra}~\cite{Basics}.


\section{Experimental aspects of parton fragmentation in the vacuum}


\tit{Colour coherence and a comparison of the fragmentation of gluon and quark jets}

\auth{Klaus Hamacher}

Coherence is an important property of gluon radiation in
the hadronic final state of inelastic interactions and a direct consequence
of the quantum nature of strong interactions. Theoretical calculations
observing coherence predict the evolution scales of jets inside of
multi-hadronic events. Typically these scales are transverse-momentum-like
and fulfill Lorentz-invariance in contrary to the often
used jet energies.
A clear demonstration of colour coherence is obtained from the measurement of the ratio
of the soft hadron multiplicity in narrow cones oriented perpendicular
to the plane of a three-jet event and the axis of a two jet event~\cite{Abdallah:2004uu,drMS}.
This ratio is predicted as the product of the colour factor ratio $C_{A}/C_{F}$
and an event scale $r_{\perp}$. The $r_{\perp}$-dependence allows
to compare the amount of coherent soft radiation due to the gluon and the 
quark colour charge. The prediction without any free parameter describes
the data very well.  A destructive $1/N_{c}^{2}$
suppressed term included in $r_{\perp}$ is required to describe the
data. The colour factor ratio when fitted to the data
is $C_{A}/C_{F}=2.211 \pm 0.014$(stat.) $\pm$ 0.053(syst.). 
The reason for this good agreement of the leading order prediction
with the data lies in the soft, large angle hadrons used as testing
probes. These hadrons due to their large wavelength are insensitive
to higher order structures and also to finite energy and leading particle
effects.\\

A precision study of the overall multiplicity of three-jet events~\cite{Abdallah:2005cy,drMS}
also relies on properly defined scales in order to describe the topology
dependence of the multiplicity. Two different choices are available
as the division of the event multiplicity in a ${\rm q\bar{q}}$ and
a gluon part is ambiguous. The most natural choice describes the data
whereas the choice minimising the amount of hadrons assigned to the
gluon jet (chosen by OPAL) fails to agree. The colour factor ratio
was fit to the slope of the scale dependence of the multiplicity yielding
$C_{A}/C_{F}={2.261}\pm0.014_{\mathrm{stat.}}\pm0.036_{\mathrm{exp.}}\pm0.052_{\mathrm{theo.}}\pm0.041_{\mathrm{clus.}}$.
This result together with a measurement of the $\beta$-function of
the thrust~\cite{Abdallah:2003xz} which agrees to the one of the coupling in NLO restrict
the gauge group of strong interactions to SU(3). From the measurement
the multiplicity of gluon-gluon colour singlet events can be extracted.
This multiplicity rises about twice as fast with the energy scale
as in the ${\rm q\bar{q}}$-case clearly demonstrating the higher
colour charge of the gluon.\\

The precise agreement of the above discussed measurements to the colour
factor expectation leaves little room for further differences in the
particle production from gluons and quarks. The measurements mainly
admit differences for leading particles. Such are indeed observed
for baryons~\cite{Abreu:2000nw}, where an excess of about a factor two (i.e. $\sim2\cdot C_{A}/C_{F}$)
is observed for leading particles. This excess is understood within
the string fragmentation model due to the higher number of possible
splitting processes into baryons. Cluster models here fail if no fundamental
splittings of gluons into diquark-anti-diquark pairs are foreseen.\\

It has been found in Monte Carlo studies that gluon fragmentation
functions obtained from three-jet events and hypothetical gluon-gluon
events disagree in specific parts of phase-space [M.~Tasevsky]. 
Therefore, the results obtained from three-jet events were quantified as ``biased''.
This distinction seems unfounded as three-jet events form the really
accessible source of data. Moreover, if the observed differences were
essential they need to be brought into accord with the QCD factorisation
theorem. It turns out, however, that the discrepancies are consequence 
of hadronisation and the necessities of the measurement process~\cite{Hamacher:2005rh}. 
The reconstruction of the parton kinematics due
to confinement must proceed by a direct identification of the jet
with the parton kinematics. However, the hadronisation transition
then leads to a smearing of the parton kinematics. Typically
this smearing vanishes $\propto1/E_{jet}$. The parton energy inferred
from the jets enters in the definition of the scaled momentum $x_{h}=p_{hadron}/p_{g/q}$
and in turn especially strongly varying fragmentation functions are
influenced by error propagation. In fact the effect is stronger for
the gluon compared to the quark fragmentation function due to the
stronger fall-off, as well as due to the somewhat increased hadronisation
smearing for gluon jets. A numerical calculation shows ratios of up
to a factor two between the smeared, measurable fragmentation function
and the underlying partonic one for gluons at high $x_{h}$ and small
$E_{jet}$. The difference detected for gluon fragmentation [M. Tasevsky] can
be fully explained by the hadronisation/measurement process. It is
important to note that similar problems unavoidably exist in any
measurement involving jets, e.g., when gluon fragmentation is measured
in ${\rm p\bar{p}\rightarrow 2~jets}$. The hadronisation smearing
cannot be properly included in an experimental unfolding as it will
depend on the fragmentation function to be measured. Still, it is
possible to use such measurements if the hadronisation smearing is
included in the phenomenological fitting procedure of the fragmentation
functions. 


\tit{ALEPH results on quark and gluon fragmentation}
 
\auth{Gerald Rudolph, for the ALEPH collaboration} 

\newcommand{\pto}{p_{\mathrm{T,out}}} 

Parton fragmentation is studied in $\epem$ annihilation into hadrons in the energy range
91 - 207~GeV using the ALEPH detector at LEP. Scaling violation in the flavour-inclusive
$x$-distribution of charged particles has been observed which is, for $x>0.1$,
qualitatively reproduced by global NLO parametrisations~\cite{scv,QCD}. The data are better
described by the QCD Monte Carlo (MC) models. The energy variation of $<N_{ch}>$ is well explained by
the QCD-MC models and by a two-parameter 3NLO calculation.\\

Inclusive $x$ distributions of a variety of identified light mesons and baryons have been
measured at the Z pole~\cite{QCD_mega,eta}. These data supplemented with $p_T$ and event
shape distributions were used to tune the free parameters of the QCD-MC programs JETSET,
ARIADNE and HERWIG~\cite{QCD_mega,CR}. The long-standing discrepancy at $\pto > 1$~GeV/c
can be slightly improved by using the new $p_T$-ordered parton shower~\cite{PYTHIA}.
The fragmentation functions of $D^{*+}$ and $B$ mesons
as well as the relative production rates of the $0^-, \: 1^-$ and
higher meson spin states in the $c$ and $b$ sector have been measured~\cite{charmBarate:1999bg,bottom,B**}.
To simulate Bose-Einstein correlations in hadronic Z decays, the
parameters $\lambda$ and $\sigma$ of the $BE_{32}$ model of PYTHIA have been determined
~\cite{BEC}.\\

The gluon and quark fragmentation functions have been measured from symmetric 3-jet
events at two scales using $b$-tagging techniques~\cite{xgq,gqstruct}. As expected, the  
gluon fragmentation function is steeper in $x$ than the quark fragmentation function.   
The failure of the QCD-MC models to describe the $x>0.4$ region of the gluon fragmentation function, first
noted in~\cite{gqstruct,OPAL}, is confirmed with high statistics in 3-jet events of general topologies with high
statistics (preliminary). The rate of neutral gluon jets with a central rapidity gap
is found to be higher than expected from JETSET or ARIADNE~\cite{CR}. The excess is
located at low invariant mass (0.8-2.2~GeV/c$^2$, preliminary), as first observed by DELPHI
~\cite{DELPHI}. The colour-reconnected versions of these generators predict a much higher
rate and thus can be excluded~\cite{CR}.

\newpage

\tit{Colour flux studies in quark and gluon fragmentation in the DELPHI experiment at LEP}
 
 \auth{Brigitte Buschbeck, for the DELPHI collaboration}

The occurrence of colour reconnection between two quark-antiquark (colour
triplet) strings~\cite{1}, Bose-Einstein correlations of their decay products
~\cite{2}, and the possible occurrence of colour octet fragmentation with the formation of
gluonic systems in the fragmentation of gluons~\cite{3} is studied in three
contributions.\\

The first contribution investigates colour reconnection directly in \epem
 $\rightarrow$ W W $\rightarrow$ ($q$1- $\bar q2$) ($q$3- $\bar q4$),
i.e. between the colour strings spanned by each W.
The second contribution investigates Bose-Einstein correlations (BEC)
between the decay products of the two different W bosons.
Both effects are determined by the space-time overlap of the two colour flux
tubes or their decay products. The data are compatible with some reconnection effects,
however with a large statistical error. Bose-Einstein correlations
between the decay products of different W's are seen by DELPHI
(significance 2.4$\sigma$ ),
however they show up weaker than predicted by a Monte Carlo
which includes full inter W correlations. Since all other LEP2
experiments do not show any signal, it is argued, that BEC may be
diminished or even absent
between the hadronisation products of different strings. This can
be a valuable tool to study the string structure in hadron-hadron and
heavy ion reactions.\\

The third contribution investigates the leading hadrons in gluon 
fragmentation in 3-jet ($e^+e^- \rightarrow q \bar q g$) events~\cite{3}.
It shows that gluon jets produce more neutral leading systems
than predicted by the LUND string models (JETSET, ARIADNE). This is
however predicted, if gluons ($g$) fragment in some cases
as a colour octet field  which is then neutralised by a $gg$ pair.
The effect is enhanced by demanding a rapidity gap and amounts at
a rapidity gap  $\Delta y$ = 1.5  of about 10 \% .
 The concentration of the excess  of the neutral systems at low invariant mass
 could be an indication for a hitherto undetected
fragmentation mode of the gluon via the formation of a gluonic system.


\tit{Fragmentation functions of quark and gluon jets as measured by OPAL}

\auth{Marek Tasevsky, for the OPAL collaboration}

\newcommand{\mr}{\mathrm}
\newcommand{\ff}{fragmentation function}
\newcommand{\ffs}{fragmentation functions}
\newcommand{\ee}{$e^+e^-$}
\newcommand{\Qj}{$Q_{\mr jet}$}
\newcommand{\xe}{$x_{\mr E}$}
\newcommand {\egstar} { E_{\mathrm g}^* }

Scaling violations of quark and gluon jet \ffs\ are studied in \ee\
annihilations at $\sqrt s =$ 91.2 and 183--209~GeV using data collected
with the OPAL detector at LEP. The scale dependence of the flavour inclusive,
$udsc$ and $b$ \ffs\ from unbiased jets is measured at $\sqrt s/2=$ 45.6 and
91.5--104.5~GeV. Biased jets are used to extract the flavour inclusive, $udsc$
and b, and gluon \ffs\ in the ranges $Q_{\mr jet}=$ 4--42, 4--105 and
4--70~GeV, respectively, where \Qj\ is the jet energy scale. Three methods
are used to extract the \ffs, namely the $b$-tag and energy-ordering methods
for biased jets, and the hemisphere method for unbiased jets. The results
obtained using these methods are found to be consistent with each other.
The $udsc$ jet results above the scale of 45.6~GeV, the gluon jet results above
30~GeV (except for the scale of 40.1~GeV), and the $b$ jet results at all scales
except 45.6~GeV represent new measurements. The results of this analysis are
compared with existing lower energy \ee\ data and with previous
results from DELPHI and OPAL. The overall consistency of the biased jet
results with the unbiased jet results suggests that \Qj\ is a generally
appropriate scale in events with a general three-jet topology.
The scaling violation is observed to be positive for lower \xe\ and
negative for higher \xe, for all the types of \ffs. The gluon jet \ff\ exhibits
stronger scaling violation than that of $udsc$ jets.\\

The bias of the procedure used to construct biased jet \ffs\ is estimated by
studying hadron level Monte Carlo generator events. In explaining the
observed differences between biased and unbiased jet results, we note the
effects of non-negligible masses of hadrons and $b$-quarks at low scales.
Due to the considerable bias found for the gluon jet \ffs\ in the region of
$x_{\mr E} > 0.6$, precautions should be taken when comparing the biased gluon
jet results with theory.
The data are compared to the predictions of NLO calculations. In a wide
range of scaled momentum \xe, all calculations satisfactorily describe the
data for the $udsc$ jet \ffs.
The description is worse and the spread between the predictions larger for the
$b$ and gluon jet \ffs, in particular in regions of very low and high \xe.
The data are also compared with predictions of three Monte Carlo models,
PYTHIA 6.125, HERWIG 6.2 and ARIADNE 4.08.
A reasonable agreement with data is observed for all models, except for high
\xe\ region with small scales ($\lesssim$ 14~GeV) in case of the $udsc$
and gluon jet \ffs.
The charged particle multiplicities of $udsc$, $b$ and inclusive hadronic
events are obtained by integrating the measured \ffs. All values are
found to be in agreement with previous measurements, where available.\\

The first experimental study to use the jet boost algorithm,
a method based on the QCD dipole model to extract properties of
unbiased gluon jets from
{\ee}$\,\rightarrow\,$$\mathrm q\overline{q}g$ events, has also
been presented. We test the jet boost algorithm using the Herwig Monte
Carlo QCD simulation program, comparing the results of this method to
those derived from unbiased gluon jets defined by hemispheres
of inclusive $gg$ events from a color singlet point source.
We find that the results of the jet boost algorithm
for the multiplicity distribution are in close
correspondence to those of the $gg$ hemispheres for jet energies
$\egstar$ larger than about 5~GeV.
For the fragmentation functions,
the results of the two methods agree to good precision
for $\egstar$$\,\gtrsim\,$14~GeV.
Therefore, the fragmentation function of unbiased gluon jets
have been measured only at two points, namely at 14.24 and 17.72~GeV.


\def\hbabar{\mbox{{\huge\bf\sl B}\hspace{-0.1em}{\LARGE\bf\sl A}\hspace{-0.03em}{\huge\bf\sl B}\hspace{-0.1em}{\LARGE\bf\sl A\hspace{-0.03em}R}}}
\def\Lbabar{\mbox{{\LARGE\sl B}\hspace{-0.15em}{\Large\sl A}\hspace{-0.07em}{\LARGE\sl B}\hspace{-0.15em}{\Large\sl A\hspace{-0.02em}R}}}
\def\lbabar{\mbox{{\large\sl B}\hspace{-0.4em} {\normalsize\sl A}\hspace{-0.03em}{\large\sl B}\hspace{-0.4em} {\normalsize\sl A\hspace{-0.02em}R}}}
\def\babar{\mbox{\slshape B\kern-0.1em{\smaller A}\kern-0.1em
    B\kern-0.1em{\smaller A\kern-0.2em R}}}


\def\B       {\ensuremath{B}\xspace}
\def\Bbar    {\kern 0.18em\overline{\kern -0.18em B}{}\xspace}
\def\Bb      {\ensuremath{\Bbar}\xspace}
\def\BB      {\ensuremath{B\Bbar}\xspace} 

\def\piz   {\ensuremath{\pi^0}\xspace}
\def\pip   {\ensuremath{\pi^+}\xspace}
\def\pim   {\ensuremath{\pi^-}\xspace}
\def\pipi  {\ensuremath{\pi^+\pi^-}\xspace}
\def\pipm  {\ensuremath{\pi^\pm}\xspace}
\def\pimp  {\ensuremath{\pi^\mp}\xspace}

\def\kaon  {\ensuremath{K}\xspace}
\def\K     {\ensuremath{K}\xspace}
\def\Kbar  {\kern 0.2em\overline{\kern -0.2em K}{}\xspace}
\def\Kp    {\ensuremath{K^+}\xspace}
\def\Km    {\ensuremath{K^-}\xspace}
\def\Kpm   {\ensuremath{K^\pm}\xspace}
\def\Kmp   {\ensuremath{K^\mp}\xspace}

\def\g     {\ensuremath{\gamma}\xspace}
\def\gaga  {\ensuremath{\gamma\gamma}\xspace}  


\def\scr{\scriptstyle}
\def\sscr{\scriptscriptstyle}

\def\ifb{\ensuremath{\,\mathrm{fb}^{-1}}}
\mathchardef\Upsilon="7107
\def\CP                {\ensuremath{C\!P}\xspace}
\newcommand{\gev}{\ensuremath{\mathrm{\,Ge\kern -0.1em V}}\xspace}

\tit{Parton fragmentation studies in \babar}

\auth{Fabio~Anulli, for the \babar\ collaboration}

The \babar\ detector~\cite{bib:babar}, operating
at the asymmetric \BF\ \pep, has been optimized for \CP violation 
studies in \B meson decays.
However, the high luminosity and the high detector performances, allow 
also  precise measurements of different aspects of strong interactions.
We present preliminary measurements of inclusive momentum spectra 
of a variety of light mesons and baryons, namely \pipm, \Kpm, $\eta$, and $p/\pbar$, 
at an energy of   $\sqrt{s}=10.54~\gev$, below the $\B\Bbar$ production threshold. 
 The  data cover  the full scaled momentum range with high precision 
(few percent relative), allowing sensitive tests of QCD calculations and 
fragmentation models.
Comparison with results from higher energies shows significant scaling
violation at high scaled momentum values, for pions and kaons, while
data for $p/\pbar$ do not present  any clear difference between 
10 and 90 \gev. 
The same spectra have been measured also in hadronic decays of 
the $\Upsilon(4S)$, providing significant inputs to model the 
inclusive properties of \B meson decays.\\

We present also a measurement of the inclusive spectra of the lightest 
charmed baryon, the $\Lambda_c^+$~\cite{bib:lambdac}.
 The scaled momentum distribution is  in agreement with the existing data
 and shows a maximum at $x_p \approx$~0.6, followed by a sharp decrease.
The total measured production rate is 
$N_{\Lambda_c} = 0.057\pm0.002$(exp) $\pm$0.015(BF)  \lambdac per hadronic event,
where the dominant uncertainty is given by the branching fraction into 
the reconstructed mode, $\lambdac\to p\Km\pip$.
The production of $\lambdac\lambdacbar$ pair has been measured at a
rate about four times higher than expected from 
the measured single \lambdac production rate; 
the event topology is not consistent with four-baryons production, while
an average of about four popcorn mesons are produced in association with 
the $\lambdac\lambdacbar$ pair.
This result is compatible with long distance baryon number conservation, 
as already pointed out by the CLEO experiment~\cite{bib:cleo-lambdac}.\\
       
The very high luminosity at the \BF\ allows the study of 
\epem annihilation into exclusive, low-multiplicity  final states.
Such processes dominate at low $\sqrt{s}$, but their cross sections 
fall much faster than $\sim 1/s$, and are several order of magnitudes 
smaller at $\sqrt{s} \sim 10~\gev$. 
Their measurement in terms of total cross section and internal structures,
 provide a  rich testing ground for QCD.
As an example, we show here the first observation of \epem annihilation in
two hadrons via two-virtual-photon-annihilation (TVPA), 
in the processes $\epem\to\rho^0\rho^0\to\pip\pim\pip\pim$ and 
$\epem\to\phi\rho^0\to\Kp\Km\pip\pim$, whose final states have 
positive charge conjugation, $C=+1$, 
and cannot be produced by a single $\gamma$~\cite{bib:rhorho}.
Annihilation of \epem pairs at low $\sqrt{s}$ is studied
in \babar\ via initial state radiation, as shown in detail in [S.~Pacetti].\\

\tit{Light-quark and spin-dependent fragmentation functions at BELLE}

\auth{Ralf Seidl, for the BELLE collaboration}

The BELLE experiment~\cite{belledetector} at the asymmetric $e^+e^-$ collider 
KEKB~\cite{kekb} has taken a large amount of data which can be used for the study 
of fragmentation functions. In particular the more than 60~fb$^{-1}$ of continuum 
data contain only $udsc$ quark-antiquark pairs and no $B$ meson decays.
In leading order, the normalized yield of inclusive identified hadrons can be directly 
related to the sum of fragmentation functions of all accessible quarks and antiquarks. 
With the statistics available systematic uncertainties will be dominating. While the 
momentum scale and smearing are very well understood the amount of misidentified 
hadrons has to be corrected. For this purpose particles where the type of the final state 
particles is known from the decay $D^*,\Lambda$ are used to obtain the particle 
detection efficiencies and fake rates. Results on the unpolarized fragmentation functions are expected soon.\\

A second type of fragmentation functions studied in BELLE treats the fragmentation of 
transversely polarized quarks. The most prominent of these fragmentation functions 
is the Collins function~\cite{collins2}. Nonzero results from a 29~fb$^{-1}$ continuum 
data sample have been published~\cite{belleprl}. The statistics of these results have been 
increased by nearly a factor of 20 as for these measurements also the data taken under the 
$\Upsilon(4S)$ could be included due to a vanishing B-meson
background in this analysis. \\

A global analysis of the published BELLE Collins asymmetries and the semi-inclusive DIS 
results from HERMES~\cite{hermesprl} and COMPASS~\cite{compassprl} containing
the Collins effect in conjunction with the quark transversity distribution were successfully 
performed by Anselmino and collaborators~\cite{alexei}. The obtained transversity distribution 
seems to be smaller than expected from bounds and the Lattice QCD predictions. However many 
aspects, such as the evolution of the Collins function are still open questions. 
Another spin dependent fragmentation function, the interference fragmentation function is also 
being studied at BELLE and results are expected soon.

\tit{Fragmentation studies at CLEO}

\auth{David Besson, for the CLEO collaboration}

Over the last two decades, CLEO has made a number of measurements testing fragmentation of 
quarks in \qqbar events vs. digluon fragmentation vs. three-gluon fragmentation at center of mass
energies around 10~GeV. Not reported in my talk are older results on suppression of decuplet 
baryons to octet baryons in inclusive \qqbar production on the continuum, inclusive pion, 
kaon and proton fractions in 3-gluon production vs. \qqbar production. Reviewed results include:
\begin{enumerate}
\item The older correlated $\Lambda_c/\bar \Lambda_c$ enhancement observed on the 
continuum~\cite{bib:cleo-lambdac}, now being redone with higher statistics by BABAR, which showed
the original enhanced baryon-antibaryon production
\item Similar $\Lambda/\bar \Lambda$ correlated production~\cite{besson2}, which showed
a similar effect, even after correcting for $c$-meson cascades. As with the previous process, 
JETSET 7.4 underestimates this enhancement by about a factor of 3.
\item Inclusive deuteron production in gluonic decays of the narrow \bbbar $\ups$
resonances~\cite{besson3}, which indicates a 3-fold enhancement of per-event
deuteron production in $ggg$ events compared to \qqbar events.
\item A comparison of particle production quark vs. gluon fragmentation, using \qqbar($\gamma$) ISR
events on the continuum to tag \qqbar events of a particular $Q^2$ vs. $gg\gamma$ events from the
narrow resonances to tag $gg$ fragmentation at the same $Q^2$. 
After confirming the large excess of baryons in $ggg$ production relative to \qqbar production, (about
double the JETSET 7.4 prediction), the photon-tagged analysis indicates only a modest
(and less momentum-dependent) baryon enhancement in $gg$ events, and better consistency with
JETSET 7.4.
\item A comparison of direct photon production in quarkonium decays. This, in turn, depends on
the color-octet and color-singlet characteristics of the recoil $gg$ states. Newer data on
the direct photon momentum spectrum from
the $\psi$ (presented first at this conference) are consistent, once accounting for two-body
radiative decays, with older data from the $\ups$~\cite{besson5}. This is
somewhat surprising, given the larger relativistic corrections in the charmonium system. We
also note that the rate of $gg\gamma/ggg$ is about a factor of 50\% larger than the value that
extrapolates directly from the $\ups$, taking into account the running of $\alpha_s$ from
the $\ups$ to the $\psi$ mass region.
\item A comparison of the average charged multiplicity in two-gluon vs. \qqbar events~\cite{besson6}, 
which showed that, at our center of mass energies, the average multiplicities are consistent with unity, to within~4\%.
\item Inclusive production of charm from the $\chi_b$ states ($J$ = 1; decaying into \qqbar, with a
nearly on-shell gluon). The level of $D^0$ production observed (approximately 10\%) is consistent
with a recent color-octet+color-singlet calculation of Braaten {\it et al.}~\cite{Bodwin:2007zf}.
\end{enumerate}

\tit{Particle production and fragmentation studies at H1}

\auth{Daniel Traynor, for the H1 Collaboration}

Results from H1 were presented for the average charged track multiplicity and for the 
normalised distribution of the scaled momentum, $x_p$, of charged final state hadrons 
as measured in deep-inelastic $e$-$p$ scattering at high $Q^2$ in the Breit frame of 
reference~\cite{h1a}. The analysis covers the range of photon virtuality
$100 < Q^2 < 20 000 ~\rm GeV^{2}$. The results are compared with $e^+e^-$ 
annihilation data and with various calculations based on perturbative QCD using different models 
of the hadronisation process.\\

The results broadly support the concept of quark fragmentation universality in $e$-$p$ 
collisions and $e^+e^-$ annihilation. A small multiplicity depletion compared to
$e^+e^-$ is observed at low $Q$ which can be attributed to higher order QCD processes 
occurring as part of the hard interaction in e-p scattering but not in $e^+e^-$
annihilation. At high $Q$ a large depletion is also observed. In the low and high $Q$ regions, 
where the comparison to $e^+e^-$ is poor, the Monte Carlo models are able to provide a 
better description of the data. The results are compared with NLO QCD calculations as implemented 
in the CYCLOPS program. All three parametrisations of the fragmentation functions used in this 
program fail to describe the scaling violations seen in the data.\\

Results from a study of mini jets in deep-inelastic electron proton scattering were presented with the
aim of finding evidence for hadronic activities in excess to those expected from the primary 
interaction~\cite{h1b}. The analysis is performed separately for the inclusive jet sample, which 
defines the towards direction, and for a dijet sample where the second jet is required to have an 
azimuthal angle larger than 140 degrees with respect to the leading jet, which defines the away 
region. The dijet sample is split into two samples which are enhanced in direct photon and resolved 
photon processes. The transverse region between the toward and away regions should be sensitive 
to additional hadronic activity. The results are compared to various QCD based models. The analysis 
covers the range $5 < Q^2 < 100 ~\rm GeV^{2}$, the leading jet is required to have a 
$p_{T,jet} > 5~\rm{GeV/c}$ as are both dijets, the mini jets must have a $p_{T}>3~\rm {GeV/c}$.\\

An overall good description of the data in the ``toward'' and ``away'' regions is given by all 
models for both the inclusive sample and the dijet sample, which proves that the models are able to 
describe the primary process. In the transverse region the predictions of the models without multiple 
interactions are generally undershooting the data for the inclusive sample and for the dijet sample 
with resolved photon events. The inclusion of multiple interactions improve the agreement 
significantly but not completely.

\tit{Particle production at ZEUS}

\auth{David H. Saxon, for the ZEUS collaboration}

Results are presented and compared to theory and models on a range of
topics using part or all of the full HERA luminosity of 0.5 pb$^{-1}$~\cite{saxon1}:
\begin{itemize}
\item {\it Charged Multiplicities and scaled momenta.} The use of the Breit
frame for comparison to $e^+ e^-$ results and the choice of scale are explained.
The benefits of using the hadronic centre of mass frame are shown. Scaled
momentum distributions are shown for $10<Q^2<40980$ ~GeV$^2$ and
compared to MC and MLLA predictions based in LEP fits. Scaling violation
is presented.
\item {\it Strange Particle Production.} $K^0, \Lambda$ and ${ \overline
\Lambda}$ distributions are presented. Compared to expectation there is an
excess of $\Lambda$'s in resolved photoproduction.
\item {\it Anti-deuteron and antiproton production.} Signals identified by
$dE/dx$ are shown. ${\overline p}$ and $p$ rates are equal whereas ${\overline d}$
production is suppressed.
\item {\it Charm fragmentation and} $F_2^{c{\overline c}}$. $D^{\star+}, D^+, D^0, D_s$
and $\Lambda^+_c$ signals are presented. Fragmentation parameters are
extracted and are mostly consistent with $e^+ e^-$ results, with an excess
of $\Lambda_c$.
\item {\it Excited charm and charm-strange mesons.} Signals are seen for a range
of mesonic states. Angular distributions indicate the likely spins.
\item {\it Baryons decaying to strange particles.} Signals are seen for
$\Lambda_c(2286), \Lambda(1520), \Xi(1320), \Sigma^\star(1385)$ and other
states.
\item $K^0_s K^0_s$ {\it Bose-Einstein correlations.} The formalism is recalled
and comparison made to $K^\pm K^\pm$. The confusion caused by $f_0(980)$
production is described.
\end{itemize}

\tit{Jets and MLLA studies at the Tevatron}

\auth{Alexandre Pronko, for the CDF collaboration}

The evolution of jets is driven by the emission of gluons at very small transverse momenta 
with respect to the jet axis. Analytical predictions are based on Next-to-Leading Log 
Approximation (NLLA)~{\cite{Basics}} calculations supplemented with the hypothesis of Local 
Parton-Hadron Duality (LPHD)~{\cite{LPHD}}. NLLA provides an analytical description of parton 
shower formation, while LPHD states that the hadronization process takes place locally, and, 
therefore, properties of hadrons and partons at the end of the parton shower are closely 
related. Detailed studies of jet fragmentation allow one to better understand the relative 
roles of perturbative parton showering and non-perturbative hadronization in shaping the 
main jet characteristics. The Tevatron data presents a unique opportunity to verify the 
validity and consistency of the NLLA+LPHD approach on a broader range of jet energies than that 
available at other machines. Overlap of the energy scales between Tevatron and $e^{+}e^{-}$
experiments allows for a direct comparison of experimental results obtained in different 
environments.\\

The inclusive momentum distributions of charged particles, $\frac{1}{N_{jets}}\frac{dN}{d\xi}$,
in jets were measured for dijet events in a wide range of dijet masses, 80-600 GeV/c$^2$~\cite{SafonovPRD}. 
The analysis was done for particles in restricted cones around the jet direction ($\theta_{C}$~=~0.28, 0.36, 0.47). 
The data were found to agree with theoretical calculations in the NLLA+LPHD framework,
in the region where NLLA is indeed applicable. A fit of the shape of the distributions 
yields the NLLA cutoff scale $Q_{eff}$~=~230$\pm$30~MeV and the ratio of charged hadrons to the 
number of partons produced in a jet $K_{LPHD}^{charged}$~=~0.56$\pm$0.10.\\

The charged particle multiplicities in quark and gluon jets, $N_{q}$ and $N_{g}$, were obtained
by comparing charged particle multiplicities in two data samples: dijet data and photon+jet data~\cite{PronkoPRL}. 
These two samples have a different quark/gluon jet content ($\sim$60$\%$ for dijet and $\sim$20$\%$ 
for $\gamma$+jet events), which allows for a measurement of the inclusive properties of gluon and quark jets. 
The evolution of measured charged particle multiplicities in gluon and quark jets with jet hardness scaling 
variable $Q$=2$E_{jet}tan({\theta}_c/2)$ agree well with trends predicted by recent 3NLLA calculations~\cite{3NLLA} 
if $Q_{eff}$ is set to 230 MeV~\cite{SafonovPRD} and normalization constant is let to be only free parameter. 
We also measured the ratio of charged particle multiplicities in gluon and quark jets, $r$~=~$N_{g}/N_{q}$. 
At $Q$~=~19 GeV, we obtain $r$~=~$N_{g}/N_{q}$~=~1.64$\pm$0.17, in agreement with 3NLLA 
calculations~\cite{3NLLA}.\\

The measurement of the two-particle momentum correlations in jets is based on dijet events with the 
invariant mass, $M_{jj}$, in the range from 66 to 563 GeV/c$^{2}$~\cite{SergoPRD}. The correlation 
function was introduced~\cite{FW} in terms of the parameter $\xi=\ln(E_{jet}/p_{hadron})$ and 
was defined as a ratio of 2- and 1-particle inclusive momentum distributions: $C(\Delta\xi_{1},
\Delta\xi_{2})=\frac{D(\xi_{1},\xi_{2})}{D(\xi_{1})D(\xi_{2})}=c_{0}+c_{1}(\Delta\xi_{1}+\Delta\xi_{2})
+c_{2}(\Delta\xi_{1}-\Delta\xi_{2})^{2}$. The results were obtained for charged particles within a 
restricted cone with an opening angle of $\theta_{c}$~=0.5 radians around the jet axis. Overall,
the data and theory show the same trends over the entire range of dijet energies. The NLLA fits~\cite{FW} 
to the evolution of correlation parameters $c_{1}$ and $c_{2}$ with jet energy scale allowed to extract 
the value of the parton shower cutoff scale $Q_{eff}$~=~$137^{+0.85}_{-0.69}$ MeV.  
The comparison to Monte-Carlo revealed that both Pythia Tune A and Herwig 6.5 reproduce correlation 
in data fairly well.\\

The measurement of the intrinsic $k_{T}$ (transverse momenta of particles with respect to jet axis) 
distributions is of particular interest because it allows to probe softer particle spectra than 
previously measured observables. The analysis was based on the same dijet events that were used in 
the described above measurement of the two-particle momentum correlations in jets. To compare 
shapes of the $k_{T}$ distributions in data with the recent theoretical calculations~\cite{PerezMachet}, 
the predictions were normalized to agree with data in the bin $-0.2<ln(k_{T})<0.0$. Within the 
validity region~\cite{PerezMachet}, the agreement between data and theory is fairly good. The inclusion 
of the higher order corrections in theoretical calculations~\cite{RedamyNMLLA} improves the agreement 
over the wider range of $ln(k_{T})$.\\



\newpage

\section{Strange and heavy quarks fragmentation: theory and experiment}

\tit{Evidence for power corrections in heavy quark fragmentation in \epem collisions}

\auth{Matteo Cacciari}\\

I report on the findings of~\cite{Cacciari:2005uk}, which compare QCD
theoretical predictions for heavy flavoured meson fragmentation
spectra in $e^+e^-$ annihilation with data from CLEO~\cite{Artuso:2004pj}, 
BELLE~\cite{Seuster:2005tr} and ALEPH~\cite{charmBarate:1999bg}. Several ingredients 
are included in the calculation: next-to-leading order initial conditions, evolution and
coefficient functions; soft-gluon resummation at next-to-leading-log
accuracy; a matching condition for the crossing of the bottom threshold
in evolution~\cite{Cacciari:2005ry} implemented at next-to-leading
order accuracy;  and important initial-state electromagnetic radiation
effects in the CLEO and BELLE data.\\

It is found that with reasonably simple choices  of a non-perturbative
correction to the fixed-order initial condition for the evolution, the
data from CLEO and BELLE can be fitted with remarkable accuracy. The
main result of the analysis is however that the fitted fragmentation
function, when evolved to LEP energies, does not reproduce fully well
the  $D^*$ fragmentation spectrum measured by ALEPH. Large
non-perturbative power corrections to the coefficient functions of the
meson spectrum are speculated to be needed in order to reconcile
CLEO/BELLE and ALEPH results. The data do not allow, however,  to
distinguish between a standard $C/Q^2$ correction with a fairly large
coefficient ($C \simeq 5$~GeV$^2$) or an unexpected linear term, $C/Q$,
with a more reasonable coefficient ($C \simeq 0.5$~GeV).

\tit{Heavy quark fragmentation with an effective coupling constant}

\auth{Gennaro Corcella}

The transition of partons into hadrons, for the time being,
cannot be calculated from first principles QCD, but is usually described in terms
of phenomenological models, containing parameters which are to be fitted to
experimental data.
Alternatively, one can model non-perturbative corrections by means of
an effective strong coupling constant, which does not exhibit the Landau pole
any longer and includes absorptive effects due to parton branching
~\cite{shirkov}.\\

The talk discusses bottom- and charm-quark
fragmentation in $e^+e^-$ annihilation,
using the effective-coupling model to include hadronization effects.
Perturbative quark production is treated in the framework of perturbative
fragmentation functions~\cite{mele},
which factorizes the production of a heavy quark
as the convolution of a coefficient function, describing the emission
off a massless parton, and a perturbative fragmentation function, associated
with the transition of a light parton into a massive quark.
The perturbative fragmentation function follows the DGLAP evolution
equations, whose solution allows one to resum the large logarithm
of the heavy-quark mass exhibited by the fixed-order calculation.
In our computation, we use NLO $\overline{\mathrm{MS}}$ coefficient
functions and NLO initial condition of the perturbative fragmentation
function. The DGLAP evolution equations are implemented in the
non-singlet sector, to NLL accuracy, and threshold resummation
is performed in the NNLL approximation, in both coefficient function and
initial condition. The effective coupling constant, evaluated to NNLO
accuracy, is the only source of non-perturbative corrections.
In order to account for the theoretical uncertainty on our predictions,
we vary the parameters entering in the perturbative computation, such as
 renormalization and factorization scales, the $b$-quark mass
and $\alpha_S(m_Z)$, within conventional ranges.\\

Following~\cite{acf}, we consider data on $B$-hadron
energy distributions at the $Z^0$ pole, collected by LEP and SLD experiments.
In particular,
the ALEPH collaboration reconstructed only $B$ mesons, while OPAL and SLD had
also a small fraction of $b$-flavoured hadrons.
We find that our calculation,
provided with the effective-coupling model, is able to give
a reasonable description of the considered data, within the experimental
and theoretical uncertainties. We also compare our model with data in
Mellin moment space, measured by the DELPHI collaboration, and obtain
fair agreement with the first five experimental moments,
within the errors.
We then investigate data on charmed hadrons from ALEPH and
$B$-factory experiments, along the lines of~\cite{cf}.
Our result is  that the effective-coupling model is able to acceptably
reproduce the ALEPH ${D^*}^+$ spectrum, while serious discrepancies are
present when comparing with $D$ and $D^*$ data from CLEO and BELLE.
In fact, other analyses had previously shown that, even with a flexible
non-perturbative fragmentation function, provided with three tunable
parameters, one was not able to reproduce at the same time both ALEPH and
$B$-factory $D$-data.
Nonetheless, we still manage to describe quite well the first ten
Mellin moments of all considered data from ALEPH, CLEO and BELLE.

\tit{Towards a model independent approach to kaon fragmentation functions}

\auth{Ekaterina Christova\footnote{Work supported by I3HP  EU network.}}

We derive different measurable quantities in kaon production in
$e^+e^-\to K+X$ and in Semi-inclusive DIS (SIDIS) $eN\to e K+X$
that measure certain Non singlet (NS) combinations of quark fragmentation functions
(FF) for kaons~\cite{we}. These quantities are obtained in a model independent way 
and hold in any order of QCD. In SIDIS, if charged $K^\pm$ are measured we have,
$\sigma_N^{K^+-K^-} \equiv \sigma_N^{K^+}-\sigma_N^{K^-}$:
\beq
&& \sigma_p^{K^+-K^-} \simeq \left( 4u_V\otimes D_u+d_V\otimes  D_d\right)^{K^+-K^-}
 \otimes (1+(\alpha_s /(2\pi))\,C_{qq})\\
&&\sigma_d^{K^+-K^-} \simeq \left( u_V+d_V\right)\otimes (1+(\alpha_s /(2\pi))\, C_{qq})\otimes
(4D_u+ D_d)^{K^+-K^-}
\eeq
that present two measurables for the two unknown FFs: $D_u^{K^+-K^-}$ and $ D_d^{K^+-K^-}$.
In particularly, this allows one to test the usually made assumption
$ D_d^{K^+-K^-}=0$. Recently very precise data
on unpolarized SIDIS  was presented by HERMES~\cite{PHD}.
If both $K^\pm$ and $K_s^0$  are measured, then the combination $K^++K^--2K_s^0$,
both in $e^+e^-$ and SIDIS kaon production, always measures the same NS combination of FFs
$(D_u-D_d)^{K^+-K^-}$,
$(\sigma ^{K^++K^--2K_s^0}\equiv \sigma ^{K^+}+\sigma ^{K^-}-2\sigma ^{K_s^0},\,
\tilde q\equiv q+\bar q)$:
\beq
 d\sigma^{K^++K^--2K_s^0}=6\sigma_0 (\hat e_u^2-\hat e_d^2)(1+\alpha_s \, C_q\otimes )
 \,D_{u-d}^{K^++K^-}\label{e+e-}\\
d\sigma_p^{K^++K^--2K_s^0}= [(4\tilde u-\tilde d)\otimes (1+\alpha_s \,C_{qq}\otimes ) +
\alpha_s \,g\otimes C_{gq}\otimes]
\,D_{u-d}^{K^++K^-}\label{SIDISp}\\
d\sigma_d^{K^++K^--2K_s^0}= [(\tilde u+\tilde d)\otimes (1+\alpha_s \,C_{qq}\otimes ) +
\alpha_s \,g\otimes C_{gq}\otimes]
\,D_{u-d}^{K^++K^-}\label{SIDISd}
\eeq
These results can be used to determine the kaon FFs. Being model independent,
they allow also one to test the parametrizations already obtained, but at different assumptions.
In addition, eqs. (\ref{e+e-})-(\ref{SIDISd}) allow to compare the FFs obtained in $e^+e^-$
at rather high $Q^2\simeq m_Z^2$, eq. (\ref{e+e-}),
with those of SIDIS at quite low $Q^2$, eqs. (\ref{SIDISp}) and (\ref{SIDISd}).
Note that as $D_{u-d}^{K^++K^-}$ is a NS, its $Q^2$
evolution does not involve any other FFs.

\def\bbr{\mbox{\slshape B\kern-0.1em{\smaller A}\kern-0.1em
    B\kern-0.1em{\smaller A\kern-0.2em R}}}
\def\ds{$D_{sJ}^*(2317)^+$}
\def\dss{$D_{sJ}(2460)^+$}
\def\dsss{$D_{sJ}^*(2860)^+$}
\def\dsx{$X(2690)^+$}

\tit{Heavy Quark Fragmentation Functions in
  $\mathrm{e}^+\mathrm{e}^-$ annihilation at $\sqrt{s}=10.6~\mathrm{GeV}$}

\auth{Rolf Seuster, for the BELLE collaboration}

Over the years, QCD was found to be a reliable theory for many
predictions. It is however restricted to regions of phase space where the
scale of the hard interaction was (much) larger than any other scale
in the particular event. In pushing the limit to its extreme, one
learns more about the theory, its failures and its
successes. One limit that can be challenged at the very successfully
operated $b$-factories is the measurement of the heavy quark
fragmentation function. At $b$-factories, charm quarks are the heaviest
quarks for which such measurements can be done, as $b$'s are produced at
threshold. For the measurement presented here, the scaled momentum
$x_p\equiv |{\vec{p}_{candidate}}|/|{\vec{p}_{max}}|$ is used throughout.
The BELLE detector~\cite{belledetector} is an experiment operated at the
asymmetric $\mathrm{e}^+\mathrm{e}^-$ collider KEKB~\cite{kekb} at a
center of mass energy of about $10.6~\mathrm{GeV}$. For this analysis, 
about $45 \cdot 10^6$ events below and $352 \cdot 10^6$ events at the
$\mathrm{\Upsilon(4S)}$ resonance were used, corresponding to
$15~\mathrm{fb^{-1}}$ and $88~\mathrm{fb^{-1}}$ of integrated
luminosity. As the contribution from $B$ hadron decay will only occur
below $x_P=0.5$, combining on- and off-resonance data above this
threshold decreases significantly the statistical uncertainty. It
allows for a very precise and detailed description of the FF 
of charmed quarks into charmed hadrons, as
shown in~\cite{Seuster:2005tr}. The difference of the on- and off-resonance data
below $x_P=0.5$ is solemnly due to the contribution from $B$ decays. The
numbers reported in~\cite{Seuster:2005tr} are in good agreement with the world
average. 
These high precision data will have to be tested against Monte Carlo
generators. Such a comparison will improve the theoretical
understanding of the underlying physics, as well as the description of
the data by the MC itself. One example for such comparisons is the determination 
of how good commonly used FFs describe the data and what
parameters of these functions to chose. Via a reweighting
technique, special MC events were generated at many parameter
points for many fragmentation functions. In general, the best
agreement was found for the Bowler and the Lund FFs.
The parameters at which these functions showed the best agreement can
be found in~\cite{Seuster:2005tr}. The commonly used Peterson {\it et al.}
fragmentation function showed by far the overall worst
performance, not agreeing with the high precision data at all. 
For details, see~\cite{Seuster:2005tr}.\\

Another example, for which such
comparisons could be made for are ratios of fragmentation functions. In
the ratio, their systematic uncertainties will partially
cancel. Whereas for the $x_p$ dependent ratio of
$\mathrm{x_p(D^{*+})/x_p(D^{+})}$, the Monte Carlo can be tuned to
give a perfect description of the data, the ratios 
$\mathrm{x_p(D_s)/x_p(D^{+})}$ and $\mathrm{\Lambda_C)/x_p(D^{+})}$
are completely mis-modelled in the Monte Carlo. No parameter will
improve the description significantly. No conclusive reason for this
discrepancy is known.

\tit{Charmed particles production in $e^+e^-\!\!\!\to c\overline{c}$ at 10.6~GeV}

\auth{Simone Pacetti, for the \bbr\ collaboration}

During the last few years many new charmed mesons and baryons
have been discovered. Some of these, especially in the charmed-strange
mesons sector, were unexpected. 
The theoretical predictions based on potential models and
the heavy quark effective theory (HQET)~\cite{hqet} fail in
describing mesons like the \ds, \dss, \dsss and \dsx. We focused
on the analyses of such states:
\begin{itemize}
\item {\boldmath \ds\ and \dss} have been observed in the decays \ds$\to D_s^+\pi^0$~\cite{ds2317}
and \dss$\to D_s^+\gamma$, $D_s^+\pi^0\gamma$, $D_s^+\pi^+\pi^-$~\cite{ds2317}. 
These analyses have been realized over the data collected by the \bbr\ experiment
in inclusive $c\overline{c}$ production at 10.6~GeV, with an integrated 
luminosity of 232~fb$^{-1}$. Masses, widths, and decay modes of these mesons do not fit the HQET predictions
based on the $c\overline{s}$ structure, hence there are speculations that
both \ds\ and \dss\ are some type of exotic state, such as four-quark or molecular
states~\cite{mol}.
\item {\boldmath \dsss\ and \dsx} have been observed in their decay into $D^{0(+)}K^+_{(S)}$,
with $D^0\to K^-\pi^+$, $K^-\pi^+\pi^0$, and $D^+\to K^-\pi^+\pi^+$~\cite{x2690}.
This analysis is realized in 240~fb$^{-1}$ of data collected by the \bbr\ experiment. 
The evidence of the lightest structure is less significant, however  
it is compatible with a similar structure observed by BELLE~\cite{belle06}.
\end{itemize}

Before the advent of $b$ factories the spectrum of charmed baryons comprised only 12 
states~\cite{pdg}. In the last years \bbr\ found 4 new charmed baryons and confirmed 3 other states
observed by BELLE.

\begin{itemize}
\item The {\boldmath $\Omega_c^*$} is a $css$ baryon with $J^P=3/2^+$, it is the first 
excited state of the $\Omega_c$ family. It has been observed by \bbr\ in the radiative decay:  
$\Omega_c^*\to\Omega_c\gamma$, in $c\overline{c}$ production with an integrated luminosity 
of 231~fb$^{-1}$~\cite{omegacs}. The mass difference with respect to the ground state $\Omega_c$
is in agreement with lattice calculations~\cite{lat} and theoretical expectations~\cite{thomegac}.
\item The {\boldmath $\Lambda_c(2940)^+$} is a $udc$ baryon, it has been observed by \bbr\ 
decaying in $D^0p$~\cite{lambdac}. It is the first observation of a charmed baryon
which decays in a light baryon, i.e.: the proton. The data for this analysis refer to an integrated
luminosity of 287~fb$^{-1}$. 
\item Two new charmed-strange baryons {\boldmath $\Xi_c(3055)^+$ and $\Xi_c(3123)^+$}, belonging to 
the $qsc$-family ($q=u,d$) have been discovered by the \bbr\ experiment, studying the 
$\Lambda_c^+K^-\pi^{+}$ system, obtained in inclusive $c\overline{c}$ production 
at 10.6~GeV with 384~fb$^{-1}$~\cite{xi}. The decay in the $\Lambda_c^+ K^-\pi^+$
final state goes trough different intermediate states, namely:
$\Xi_c(3055)^+\to \Sigma_c(2455)^{++}K^-\to \Lambda_c^+\pi^+ K^-$
and $\Xi_c(3123)^+\to \Sigma_c(2520)^{++}K^-\to \Lambda_c^+\pi^+ K^-$.
No neutral partners have been observed.
\end{itemize}

\tit{ISR Physics at \bbr}

\auth{Simone Pacetti, for the \bbr\ collaboration }

Using the initial state radiation (ISR) technique in fixed center-of-mass
energy machines, like the flavour factories, we may mimic a typical
$e^+e^-$ annihilation experiment with an energy scan in the center of mass.
At the Born approximation, in the differential cross section for processes like
$e^+e^-\to \gamma_{\rm ISR}+$hadrons, the function describing the radiation
of a photon ($\gamma_{\rm ISR}$) by one of the initial leptons and
the hadronic cross section factorizes. This allows us to perform precise
measurements of hadronic cross sections at low energy (from threshold up to
4-5~GeV). The knowledge of 
$R=\sigma(e^+e^-\to\rm hadrons)/\sigma(e^+e^-\to\mu^+\mu^-)$
is crucial to compute the hadronic contributions to the muon anomalous 
magnetic moment~\cite{mu} (low-energy values), and to the running
electromagnetic coupling $\alpha_{\rm QED}$~\cite{alpha} (high-energy values).
In addition we perform a better parameter determination for many
$\rho$-, $\omega$-, and $\phi$-excited states ($J^{PC}=1^{--}$).
\begin{itemize}
\item The light-meson final states measured by the \bbr\ experiment 
     in the 1-3~GeV region are:
    $\pi^+\pi^-\pi^0$, $2(\pi^+\pi^-)$, $K^+K^-\pi^+\pi^-$~\cite{3pi},
    $3(\pi^+\pi^-)$, $2(\pi^+\pi^-\pi^0)$, $K^+K^-2(\pi^+\pi^-)$~\cite{6pi}.
\item A new narrow ($\Gamma\sim$ 60 MeV) vector meson called $X(2175)$ has been discovered decaying 
      in $\phi f_0(980)$~\cite{x2175} and confirmed in the $\phi\eta$ channel~\cite{x2175_2,noi}.
\item We have analyzed the reaction $e^+e^-\to KK\pi$~\cite{noi} in two samples: $K^+K^-\pi^0$ and
      $K_SK^\pm\pi^\mp$. The study of the $K_SK^\pm\pi^\mp$ asymmetric Dalitz plot 
      allowed, for the first time, the identification of the two isospin components and their 
      relative phase. We performed a global fit, exploiting five different data sets, to 
      describe the total  $KK\pi$ cross section in terms of intermediate resonances. We determined
      parameters for $\phi$- and $\rho$-excited states,
      with a special care to the OZI-suppressed $\phi\pi^0$. 
      The $\phi\pi^0$ cross section, with a peak value of $\sim0.2$ nb, shows
      two resonant structures: a broad bump at $\sim1.6$~GeV, compatible with a $\rho$ recurrence~\cite{pdg}
      and a narrower resonance at $\sim1.9$~GeV, whose parameters are in agreement with those of
      the ``dip'' observed in multi-hadronic final states~\cite{dip}.
\item The \bbr\ experiment measured with unprecedented accuracy the differential
      cross section for $e^+e^-\to p\overline{p}$~\cite{pp}. In particular, by studying
      the angular distribution, \bbr\ provides also a time-like measurement for the ratio
      $|G_E^p/G_M^p|$ of the electric and magnetic proton form factors. We observed, 
      for the first time, a displacement from the ``scaling behavior'':  $|G_E^p/G_M^p|\simeq1$,
      by measuring $|G_E^p|>|G_M^p|$ near threshold. The angular distribution provides
      also a powerful tool to estimate the two-photon exchange contribution in 
      the annihilation $e^+e^-\to p\overline{p}$. Using the \bbr\ data this 
      contribution turns out to be compatible with zero~\cite{egle}.
      In addition, the $p\overline{p}$ 
      cross section has two very interesting features: a non-vanishing threshold
      value, which is however expected as a consequence of the $p\overline{p}$\, Coulomb interaction, 
      and the presence of structures (``steps''), not yet understood, at $\sim 2.2$ and 
      $\sim 3.0$~GeV.
\item Other baryon-antibaryon final states like: 
      $\Lambda\overline{\Lambda}$, $\Sigma^0\overline{\Sigma^0}$ and 
      $\Lambda\overline{\Sigma^0}$~\cite{ll},
      have been measured by \bbr\ with the ISR technique. 
      In all these cases the cross section data have finite threshold values, 
      as in the $p\overline{p}$ cross section, which are not
      well understood. Indeed, having always neutral baryons, no Coulomb 
      interaction effect can be used to cancel the vanishing of the phase space.
\end{itemize}


\tit{Study of charm fragmentation in DIS at H1}

\auth{Zuzana R\'urikov\'a, for the H1 collaboration}

The process of charm fragmentation is studied in deep-inelastic 
$D^{*\pm}$ meson production as measured by the H1 detector
at HERA~\cite{preliminary}. Two different methods are used for the 
definition of the observable approximating the fraction of the 4-momentum of
the  $D^{*\pm}$ meson with respect to the charm quark.
The $c$-quark momentum is approximated in the
$\gamma^{*}p$-rest-frame either by the momentum of the jet including
the $D^{*\pm}$ meson or by the momentum of a suitably defined hemisphere which
includes the $D^{*\pm}$ meson.\\

The visible phase space of this analysis is given by cuts on the photon
virtuality $2<Q^{2}<100$~GeV$^{2}$, event inelasticity $0.05<y<0.7$ and
$D^{*\pm}$ meson phase space $1.5<p_{\rm T}(D^{*\pm})<15$~GeV/c,
$|\eta(D^{*\pm})|<1.5$. In addition a $D^{*\pm}$ jet with $E_{\rm T}^{*}>3$~GeV
in $\gamma^{*}p$-rest-frame is required in order to have hard scale for the events.
The fractional energy distributions of $D^{*}$ mesons, corrected for detector
effects, are used to extract parameters for the Kartvelishvili and Peterson
fragmentation functions within the framework of leading order + parton shower
QCD models RAPGAP 3.1 and CASCADE 1.2 with string fragmentation and particle
decays as implemented in PYTHIA, as well as for the next-to-leading order QCD
calculation in the fixed flavour scheme HVQDIS~\cite{hvqdis}.\\

The fragmentation parameters extracted using both observables are in good
agreement with each other. Furthermore the value of Peterson parameter
$\varepsilon$, extracted for the PYTHIA parameter setting~\cite{ALEPH-steering},
is in agreement the ALEPH tuned value of $\varepsilon= 0.04$, supporting the
hypothesis of fragmentation universality between $e^{+}e^{-}$ and $e$-$p$.
Finally, the fragmentation of charm produced close to the kinematic threshold
is studied. The description of this sample by the QCD models and NLO
calculation is not good, and the extracted fragmentation parameters are
significantly different from the parameters fitted to the nominal
sample with a hard scale. This can be interpreted as a hint of inadequacies of
the QCD models and possibly of insufficient flexibility of the simple
parametrisations used for the fragmentation function in the phase space
region close to kinematic threshold.

\newcommand{\ksks}{$K_S^0 K_S^0$\xspace}
\newcommand{\Ks}{$K_S^0$\xspace}
\newcommand{\ffa}{$f_{2}(1270)/a_{2}^{0}(1320)$\xspace}
\newcommand{\ffb}{$f_{2}^{'}(1525)$\xspace}
\newcommand{\ffc}{$f_{0}(1710)$\xspace}

\tit{Inclusive \ksks resonance production in $e$-$p$ collisions at HERA}

\auth{Changyi Zhou, for the ZEUS Collaboration}

The lightest glueball is predicted to have $J^{PC}=0^{++}$~\cite{zhou1,zhou2}
and a mass in the range 1450--1750 MeV/c$^2$. Thus, it can mix with $q\overline{q}$ states
from the scalar meson nonet, which have $I=0$ and similar masses.
In the literature, the state \ffc is  frequently considered to be a state
with a possible glueball or tetra-quark composition (see for example the reviews
~\cite{zhou3,zhou4}).\\

Inclusive \ksks production in $e$-$p$ collisions dominated by photoproduction with
exchanged photon virtuality, $Q^2$,  below 1 $\gev^2$, at HERA has been studied 
with the ZEUS detector using an integrated luminosity of 0.5~fb$^{-1}$ from HERA-I and HERA-II 
combined. \Ks mesons were identified through the charged-decay mode, $K_s^0 \to\pi^{+}\pi^{-}$, 
and reconstructed with various track quality cuts for selection. The \ksks invariant mass distribution 
was reconstructed by combining two \Ks candidates.
 Enhancements in the mass spectrum have been observed attributed to the production of 
$f_2$(1270)/$a_2^0$(1320), $f_2'$(1525) and $f_0$(1710). The three states $f_{2}(1270)$, $a_{2}^{0}(1320)$
and \ffb are all of $J^P = 2^+$ and so their interference is seen in the total
cross-section. The intensity is the modulus-squared of the sum of these
three amplitudes plus the incoherent addition of \ffc and the
non-resonant background. The amplitudes for $f_{2}(1270)$, $a_{2}^{0}(1320)$
and \ffb production were fixed at the SU(3) ratios of $5:-3:2$ as expected for production via an
electromagnetic process~\cite{zhou5,zhou6}.
Very competitive measurements on peak position and width for \ffb and \ffc are done with interference 
fit and the overall fit describes the data very well. Complete systematic checks have been performed. 
The final values with statistical and systematical uncertainties were compared well with the PDG values~\cite{pdg}.


\tit{RHIC constraints on fragmentation functions for strange hadrons in $p$-$p$ collisions
at $\sqrt{s}=200$~GeV}

\auth{Mark Heinz, for the STAR collaboration}

\newcommand{\hmin}{\ensuremath{h^{-}}\xspace}
\newcommand{\lam}{\ensuremath{\Lambda} \xspace}
\newcommand{\lams}{\ensuremath{\Lambda}s\xspace}
\newcommand{\alam}{\ensuremath{\overline{\Lambda}}\xspace}
\newcommand{\alams}{\ensuremath{\overline{\Lambda}s}\xspace}
\newcommand{\pbars}{\ensuremath{\overline{p}s}\xspace}
\newcommand{\pbarp}{\ensuremath{\overline{p}+p}\xspace}
\newcommand{\kp}{\ensuremath{K^+}\xspace}
\newcommand{\km}{\ensuremath{K^-}\xspace}
\newcommand{\pion}{\ensuremath{\pi}\xspace}
\newcommand{\Kpi}{\mbox{\ensuremath{<\!\!K\!\!>/<\!\!\pi\!\!>}}\xspace}
\newcommand{\Kpip}{\mbox{\ensuremath{<\!\!K^{+}\!\!>/<\!\!\pi^{+}\!\!>}}\xspace}
\newcommand{\Kpim}{\mbox{\ensuremath{<\!\!K^{-}\!\!>/<\!\!\pi^{-}\!\!>}}\xspace}
\newcommand{\kz}{\ensuremath{\mathrm{K^{0}_{S}}}\xspace}
\newcommand{\rap}{\ensuremath{{\bf y}}\xspace}
\newcommand{\midrap}{\ensuremath{{|{\bf y}|<0.5}}\xspace}
\newcommand{\mt}{\ensuremath{m_{T}}\xspace}
\newcommand{\XI}{\ensuremath{\Xi^{-}}\xspace}
\newcommand{\axi}{\ensuremath{\overline{\Xi}^{+}}\xspace}
\newcommand{\ppbar}{\ensuremath{p+\bar{p}}\xspace}
\newcommand{\sqsRhic}{\ensuremath{\sqrt{s}=200}\xspace}
\newcommand{\sqsSps}{\ensuremath{\sqrt{s}=630}\xspace}

Perturbative QCD has proven successful in describing inclusive
hadron production in elementary collisions. Within the theory's
range of applicability, calculations at next-to-leading order (NLO)
have produced accurate calculations for the \pt 
spectra of charged hadrons in different collision systems and energies, 
which has lead to claims of universality of the underlying fragmentation 
functions (FFs)~\cite{KKP}. With the new mid-rapidity $p$-$p$ data at \sqsRhic
GeV collected at RHIC, comparisons to pQCD can now be extended to
identified strange baryons and mesons~\cite{Abelev:2006cs}.
In the last 5 years significant improvements have been made in the
field of NLO FF studies. Several groups have updated their parameterizations 
to include not just \ee\ but also (SI)DIS and now $p$-$p$ data, thus improving the
constraints on the FF parameters. In particular the
gluon-to-hadron FF was never well constrained by \ee\ and DIS data
due to the low probability of gluon-jet production. Conversely, at RHIC we
are probing a very gluon-jet rich environment and the majority of
final states are produced by gluons. In particular baryons are
dominated by gluon fragmentation. According to AKK,
90\% (10\%) of the protons below \pt~=~10~GeV/c are produced from gluons (quarks).
On the other hand, pions are produced in equal parts by gluons and quarks.\\

Experimentally this fact can be exploited to measure the
baryon-to-meson ratios and compare them to LO/NLO calculations.
Results by the STAR collaboration confirm that LO models are not
able to accurately describe this ratio~\cite{Adams:2006nd,Abelev:2006cs} 
and underpredict the amount of
baryons at low \pt. For strange baryon production ($\Lambda$) the first NLO
predictions equally fail to describe the data~\cite{DSV}. In an
attempt to solve this shortcoming, one group (AKK)~\cite{AKK}
readjusted the gluon FF to fit the data, 
finding $D_{g}^{\lam}= D_{g}^{p}/3$. Another group (DSS) has for
the first time performed separate fits to protons and anti-protons
based on STAR data~\cite{DSS2}, and provided new FFs for pions,
kaons and charged hadrons from a global analysis~\cite{DSS1}.
Another subject which did not receive sufficient attention up until
recently are error estimations on the FF parameterizations. One
group (HKNS)~\cite{HKNS} has now published a global (Hessian) error
analysis for charged hadrons and finds that the 4 most ubiquitous FF
agree within errors. However this analysis does not include $p$-$p$
~data.
The STAR experiment has also measured identified particle spectra in
terms of \mt~=~$\sqrt{\pt^{2}+m^{2}}$ as motivated by earlier
studies~\cite{Mtscaling}. After arbitrary normalization of
the different particle species, this representation clearly shows
different shapes of the spectra for the baryons and mesons for $\mt >$~2~GeV/c$^2$. 
Interestingly, a simulation of the same observable with PYTHIA (LO pQCD) 
shows a similar feature when selecting gluon-jet enriched events~\cite{PYTHIA}.
Finally, efforts at RHIC are now being focused in the extraction of
medium modified FFs. Several approaches are being pursued in
parallel, such as di-hadron correlations, 3-particle correlations
and also full jet reconstruction.

\newpage

\section{Parton fragmentation in cold nuclear matter: theory and experiment}

\tit{$p_T$-broadening in nuclear DIS}

\auth{Hans-J\"urgen Pirner}

My work deals with modifications of fragmentation processes in cold nuclear matter and in 
the hot quark gluon plasma. In the first part a phenomenological analysis of deep-inelastic hadron 
production in nuclei is performed. The main theoretical picture is a three-stage model which includes 
partonic propagation, prehadron formation and absorption, and finally hadron formation.
 
This model has been successfully applied to describe the HERMES data presented 
in [E.~Aschenauer]. The multiplicity ratios of $\pi, K,p,\bar p$ produced in
$e$-Ne, He, Kr, Xe collisions  as a function of the momentum fraction of the hadrons $z_h$, 
and the photon energy $\nu$ can be modelled by a prehadron-nucleon absorptive cross section 
of one third of the hadron cross section. No $Q^2$-dependence of the multiplicity ratios is 
calculated as observed in the data.\\

The second part emphasizes the $p_T$-broadening in the hadron spectra as a tool 
to monitor the path length of the quark in the nucleus. The meson-nucleon interactions have 
a small elastic cross sections relative to the total cross section, and can be 
neglected for $p_T$-broadening. Again the calculated differential  $\Delta p^2_T$ broadening 
as function of  $z_h,\nu,Q^2$  are compared with data. The observed log($Q^2$) dependence 
indicates the interleaving of parton radiation  and parton-medium scattering in the evolution process.\\

Both effects can be described with a new equation where a splitting kernel and a 
scattering kernel are present in the evolution equation. Consequences of this equation  for the 
jet profiles in LHC experiment have been shown and further details will be given in [S. Domdey].


\tit{Hadronization of pions and kaons from nuclei using DIS}

\auth{Kenneth Hicks, for the CLAS Collaboration}

The CLAS experiment {\sc eg2} was run with a variety of nuclear 
targets using a 5.5~GeV electron beam from the 
continuous electron beam accelerator facility (CEBAF) at Jefferson Lab.
The goal of this experiment is to measure observables related to 
the propagation of a quark (struck by the virtual photon from 
deep inelastic scattering) through cold nuclear matter. These 
results could be contrasted with quark propagation through hot 
QCD matter, as in the 
quark-gluon plasma expected to be formed in 
relativistic heavy ion collisions (RHIC). Quarks from RHIC are 
tagged by back-to-back ``jets'' of high energy hadrons formed in 
hard parton collisions.  In contrast with RHIC results, where 
the quark propagation is severely damped by passage through hot 
QCD matter, the preliminary DIS results suggest that the struck quark 
propagates relatively freely (compared with fully formed mesons) 
through cold nuclear matter.\\

The measured quantities of the DIS measurements are the squared 
four-momentum transfer ($Q^2$) and energy transfer ($\nu$) from 
the scattered electron and the energy fraction ($z = E_h/\nu$) 
and transverse momentum ($p_T$) of the leading hadron.  By comparing 
the $z$ and $p_T$ distributions of pions and kaons from a deuterium 
target with that from various nuclear targets, the quark propagation 
and formation into a hadron can be inferred.  Theoretical predictions 
in a given model provide guidance for interpretation of the results. 
Ratios of the number of hadrons detected, normalized to the number 
of DIS events in a given kinematic bin of $Q^2$ and $\nu$, for a 
nuclear target divided by the same quantities for a deuterium target 
are one quantitative measure of the effects of propagation through 
cold nuclear matter.  The subtraction of the average value of $p_T^2$ 
seen for a nuclear target minus that for a deuterium target is 
another measure which, according to theoretical models, is sensitive 
to gluonic radiation by the quark before it forms into a hadron. 
By comparing these two quantitative measures with theoretical 
calculations over a variety of kinematics, then models of hadronization 
can be tested.  The HERMES experiment has already published the 
nuclear attenuation ratios for several types of hadrons at DIS 
kinematics in the range $8<\nu<23$~GeV.  The CLAS data are for a 
lower range, with $2.6<\nu<4.3$~GeV corresponding to shorter 
formation times.  At HERMES kinematics, calculations suggest that 
the quark propagates through the full nucleus, followed by hadron 
formation outside of the nuclear radius.  At CLAS, we expect to 
see hadron formation inside the radius of larger nuclei.\\

Preliminary results for positive pions produced at CLAS for the 
{\sc eg2} experiment indicate a plateau for the difference $\Delta p_T^2$ 
as the nuclear size increases (targets C to Pb).  This suggests 
that, at these kinematics, the average hadron formation length 
is shorter than the radius of Pb.  In contrast, nuclear attenuation 
ratios cannot be explained by theoretical models unless some 
finite propagation of a ``pre-hadron'' (with a smaller cross section 
than fully-formed hadrons) is assumed.  Attenuation ratios for 
neutral kaons appear to be smaller than those for $\pi^+$ particles, 
but results are still preliminary.
In conclusion, the CLAS data provide a new kinematic window for 
hadronization at the expected length scale near to the radius of 
heavy nuclei.  Statistical uncertainties are small due to an 
experimental luminosity of about 100 more than for the HERMES data. 
Both $\pi^+$ and $K_s^0$ nuclear attenuation ratios and $\Delta p_T^2$ 
are expected to be submitted for publication before the end of 2008.

\tit{Characterising hadronization with nuclear DIS} 

\auth{Valeria Muccifora}

\newcommand{\open}{{<\kern -0.3 em{\scriptscriptstyle )}}}
A review on the recent progress in the study of the parton propagation, interaction and hadronization in 
{\it cold} nuclear matter has been presented. It has been pointed out that the cleanest environment to 
address nuclear modification of hadron production is the nuclear Deep Inelastic Scattering:it allows to 
experimentally control many kinematic variables; the nuclear medium (i.e., the nucleus itself) is well 
known; the hadron multiplicity in the final state is low, allowing for precise measurements. Moreover, 
the nucleons act as femtometer-scale detectors of the hadronizing quark, allowing to study its space-time 
evolution into the observed hadron.\\ 

The experimental highlights from deep inelastic lepton nuclear scattering have been shown, in particular 
results from the HERMES experiment have been discussed.
At HERMES, nuclear semi-inclusive deep-inelastic scattering is used to study the quark propagation and 
hadronization  via single hadron multiplicity ratio ~\cite{hermess}, double hadron production~\cite{hermesd}
and broadening of the hadron transverse momentum $p_T~$~\cite{ptbroad}. While hadron multiplicity ratio 
is sensitive to all stages of the hadronization process, i.e. to the production time $t_p$ of the colorless 
$q\bar{q}$ system and to the formation time $t_f$ of the final hadron, the $p_T$-broadening has been shown 
to be sensitive to  $t_p$ because, once the $q\bar{q}$ is formed (t $>$ $t_p$) , no further broadening occurs 
as inelastic interaction is suppressed.\\ 

The most recent theoretical frameworks for describing the interaction of energetic partons and space-time 
evolution of the hadronization process in nuclear DIS have been discussed. In particular models based mainly
on parton energy loss effect~\cite{wang,arleo} have been compared with models based mainly on the 
interaction of the $q\bar{q}$ system in the medium and that are sensitive to the time evolution of the 
hadronization in the medium~\cite{kopel,gellmeister}. 
It has been pointed out the importance to study parton propagation and interaction in {\it cold} nuclear 
matter for the interpretation of data in A-A collisions. In particular the connections between the kinematic 
in DIS and in nucleon-nucleon collisions have been discussed and the equivalent phase spaces covered by 
different experiments have been compared~\cite{acca2}.

\tit{Parton propagation and hadron formation: present status, future prospects }

\auth{Will Brooks}

The process of hadronization is unique to QCD, reflecting the non-Abelian nature of the 
strong interaction. For pragmatic descriptions of scattering phenomena, hadronization is 
captured by the empirically determined fragmentation functions. Until recently, study of 
the parton propagation and hadron formation processes was limited to asymptotic properties 
of the final state, such as hadron multiplicities; any microscopic information on the space-time 
development of the process on the femtometer scale was lost.
The microscopic space-time development of hadronization can now be accessed using the nuclear 
medium as an analyzer. The interactions with the nuclear medium can be used to refine the 
understanding of the mechanisms at work at $10^{-15}$ m, yielding new insights into such 
fundamental processes as gluon emission by a quark and the enforcement of color confinement 
through hadron formation. Extraction of confinement quantities such as the lifetime of the 
deconfined quark now appears to be possible. Cross comparison of new data in a variety of 
reactions will potentially provide an exciting, coherent picture, although several important points 
concerning the interaction with the medium still need to be clarified. Foremost among these for 
the semi-inclusive DIS data is the question of whether the observed hadron attenuation is primarily 
due to prehadron absorption, or due to medium-simulated gluon emission. Further, the 
extent to which information derived from cold nuclear matter can be used to interpret data from 
hot nuclear matter continues to be a subject of discussion. In particular, medium-induced gluon 
emission is broadly believed to be involved in both cases, offering the possibility of constraining the 
interpretation of the high-energy heavy-ion data using the well-understood properties of cold nuclear matter.\\

In addition to semi-inclusive DIS, Drell-Yan data offer another venue in which to study propagating 
quarks in cold nuclear matter, without the additional complication of hadron formation in the final 
state. In this talk, transverse momentum broadening in Drell-Yan and semi-inclusive DIS were 
compared, with significant differences between the Drell-Yan data from HERMES [E.~Aschenauer,V.~Muccifora] 
and Jefferson Lab [K. Hicks]. While several possibly 
relevant aspects of these differences are understood, a comprehensive picture awaits further theoretical 
development. It should be noted that the extremely high precision of both of the latter two, quite new, 
data sets is unprecedented and these data should stimulate a new wave of theoretical activity. The JLab 
data for positive pions comes as a fully three-dimensional function of $Q^2$, $\nu$, and $z$, while the 
HERMES results include both $\pi^\pm$ and an exploratory look at kaons. Further, the increase of 
$\Delta p_T^2$ in the JLab data 
appears to saturate at high nuclear mass. This is due 
to the lifetime of the deconfined quark being shorter than the dimensions of the heaviest nuclei, 
and a quantitative extraction of this lifetime within model assumptions is clearly feasible.
An inter-comparison of the HERMES and JLab results for hadron attenuation shows a good consistency. 
The landmark HERMES data set has more than half a dozen different hadrons, 
extending to much higher $\nu$ than the JLab data, and thus provides the best constraint on the 
$\nu$ and flavor dependence. The JLab data, while limited thus far to pions and the first $K^0$ study, 
nonetheless offers two orders of magnitude more luminosity, thus permitting three dimensional 
binning of the multiplicity ratio for, e.g., $\pi^+$. Studies of these data thus far do reveal 
systematic variations of the multiplicity ratio in $Q^2$, $\nu$, $p_T^2$, and $z$, with the largest 
variation in $z$ and in $p_T^2$, in agreement with HERMES.\\ 

While further studies of the existing data are ongoing, the next significant advance in the field will 
come with the 12-GeV Jlab upgrade. The upgrade of CLAS will permit yet another 
order of magnitude increase in luminosity, adding baryon fragmentation studies, as well as a broader 
range in $\nu$, $Q^2$, and $z$. The increased luminosity allows access to formation of rarer and 
heavier mesons such as the $\phi$. In total, eleven mesons and eight baryons are included in the 
existing plans with an 11 GeV $e^-$ beam.


\section{Parton fragmentation in  hot\hspace{0.1cm}\&\hspace{0.1cm}dense QCD matter:\hspace{0.1cm}theory\hspace{0.1cm}\&\hspace{0.1cm}experiment}

\tit{ Medium-evolved fragmentation functions}

\auth{Carlos Salgado}

Medium-induced gluon radiation is usually identified as the dominant dynamical mechanism underling 
the {\it jet quenching} phenomenon observed in heavy-ion collisions (see e.g.~\cite{CasalderreySolana:2007pr} 
for a recent review on the calculation of the spectrum and the corresponding phenomenological implementation). 
In its present implementation, multiple medium-induced gluon emissions are assumed to be independent, 
leading, in the eikonal approximation, to a Poisson distribution~\cite{Salgado:2003gb}.
In~\cite{Armesto:2007dt} we have introduced a medium term in the splitting probabilities
\begin{equation}
P^{\rm tot}(z)= P^{\rm vac}(z)+\Delta P(z,t),
\label{eq:medsplit}
\end{equation}
so that both medium and vacuum contributions are included on the same footing in a DGLAP approach. 
In eq.~(\ref{eq:medsplit}), $\Delta P(z,t)$ is taken directly from the medium-induced gluon radiation spectrum %
\begin{equation}
\Delta P(z,t)\simeq \frac{2 \pi  t}{\alpha_s}\,
\frac{dI^{\rm med}}{dzdt} ,
\label{medsplit}
\end{equation}
computed in the multiple scattering approximation~\cite{Wiedemann:2000za}.
The improvements include energy-momentum conservation at each individual splitting, medium-modified 
virtuality evolution and a coherent implementation of vacuum and medium splitting probabilities. 
Noticeably, the usual formalism is recovered when the virtuality and the energy of the parton are very large. 
This leads to a similar description of the suppression observed in heavy-ion collisions with values of the 
transport coefficient of the same order as those obtained using the {\it quenching weights}.
In a previous publication~\cite{Polosa:2006hb} it has been found that this formalism would 
lead to non-trivial angular dependences of the jet profiles when kinematic constrains are imposed. Similar 
structures were found at RHIC in two-particle correlations in the away side~\cite{:2008cq}.

\tit{Hadronic composition as a characteristics of jet quenching at the LHC}

\auth{Sebastian Sapeta}

There are several mechanisms which may lead to modification of the hadronic composition of jets 
(hadrochemistry), {\it e.g.} exchange of quantum numbers (color, flavor, baryon number) between 
projectile and the medium, recombination of partons from the jet with partons from the medium as well as recoil effects.
All the above may require serious modeling.  Instead, we consider the framework which takes into account only 
the exchange of momentum between the medium and the developing partonic cascade.\\

To calculate single particle distributions of identified hadrons we use the framework of Modified Leading 
Logarithmic Approximation (MLLA)~\cite{Dokshitzer}. This perturbative approach combined with the 
hypothesis of Local Parton-Hadron Duality (LPHD) was shown to reproduce correctly
not only the distributions of all charged particles but also the spectra of identified hadrons such as pions, 
kaons and protons~\cite{LPHD,Azimov:1985by}. To model the medium-modification of jets 
we supplement the above formalism by the formulation of parton energy loss proposed in~\cite{Borghini:2005em}. 
The effects of medium-induced gluon radiation are introduced by enhancing the singular parts of the 
leading order splitting functions by the factor $1+f_{\rm med}$. This leads to the softening of hadron 
spectra. The model accounts for the nuclear modification factor at RHIC when $f_{\rm med}$ is of the order of 1.\\

Our main result are  the ratios of $K^{\pm}/\pi^{\pm}$ and $p^{\pm}/\pi^{\pm}$~\cite{Sapeta:2007ad}. 
We observe a significant difference between the ratios in vacuum and medium jets. According to our analysis,
in the presence of medium the ratios increase up to 50\%  for kaons and 100\% for protons. This enhancement 
seems to be a generic feature for radiative energy loss jet quenching models. In addition we observe only mild 
dependence on the the jet energy and the jet opening angle. The medium modified hadron spectra may be also 
superimposed on the high multiplicity environment of heavy ion collisions expected at the LHC. We have checked 
that due to the characteristically different hadrochemistry of the jet and the background our result concerning the ratios
remains virtually unchanged. Hence, we conclude that the hadrochemical composition of jets may be
very fragile to the medium effects and could be used as an additional handle to study microscopic mechanisms 
underlying the jet quenching phenomenon.

\tit{QCD evolution of jets in the quark-gluon plasma}

\auth{Svend Domdey}

Based on our recent paper~\cite{DPZSIR}, I briefly discuss the
effect of scattering for jet quenching of gluonic fragmentation
functions in the quark-gluon plasma. In contrast to the common
discussion~\cite{Wiedemann:2000za,Salgado:2003gb,jetquench} 
which treats medium-induced gluon emission as the main mechanism, we find that energy loss from scattering
will also contribute significantly.
The basic idea given in the talk is as follows: The DGLAP equation
is build from the probability for splitting due to change of
virtuality. The time scale for this process is
$\tau_{split}=E/Q^2$. We extend this equation to contain also
scatterings with gluons from the medium by constructing a
``scattering probability'' from the ratio of the time scale for
splitting and the scattering mean free path $\tau_{scat}=1/(\sigma
n_g)$, i.e. the number of scatterings which can happen between two
splittings. Similar to splitting the medium-modified evolution
equation contains a gain and a loss term.\\

We discuss two slightly different approaches: One is more suitable
for analytical calculations, the other one for implementation in a
Monte Carlo study. Both methods give very similar results.
Even with a perturbative cross section, we find a suppression of
the fragmentation function in medium relative to vacuum which is
increasing with energy fraction $x$. For $Q$ =100~GeV and a plasma
temperature of $T$ = 500~MeV (which may be relevant for LHC) we find
a suppression of 10\% at $x\simeq 0.1$, but a suppression of
factor 2 and more for large $x$. For RHIC conditions, this
suppression is more moderate.

\tit{Jet reconstruction performance in the ALICE experiment}

\auth{Magali Estienne}

Under extreme conditions of pressure or temperature, nuclear matter is
expected to undergo a phase transition to a deconfined medium, the
quark gluon plasma (QGP). Such an atypical state of matter can be
defined as a very high density medium in which free partons strongly
interact. Experimentally, it is possible to re-create the conditions
necessary to its formation by colliding heavy ions at
ultra-relativistic velocity. It is part of the experimental program
which will be performed at the Large Hadron Collider at CERN. ALICE,
one of the four LHC experiments, is dedicated to the study of the QGP
properties.\\

In QCD, jets are defined as cascades of consecutive emissions of
partons initiated by partons from an initial hard scattering. This
process is followed by the fragmentation of partons into clusters of
particles emitted in some opposite collimated directions. In heavy ion
collisions, the scene of parton fragmentation is modified from
vacuum to a dense medium. 
Before they fragment, partons are
expected to lose energy through collisional energy loss and medium
induced gluon radiation, dominant process in a QGP, also called ``jet quenching''~\cite{jetquench}. 
Direct consequences of this quenching effect are
a modification of the jet shape as well as of the parton fragmentation
function associated~\cite{Borghini:2005em}. The ALICE experiment is planning to
measure these modifications thanks to its excellent tracking
capabilities down to momenta of order of $\Lambda_{QCD}$ and up to 100~GeV/c 
and thanks to its large PID capabilities~\cite{estienne3}.\\

Different types of jet finding algorithms (cone and k$_T$)
are under study in ALICE for both $p$-$p$ and Pb-Pb collisions~\cite{estienne4,estienne5}. 
Because of the large underlying background in a jet cone of radius $R = \sqrt{\Delta \eta^{2}+\Delta
\phi^{2}}$~=~1 as well as its large fluctuations in the LHC heavy ion environment, 
some studies have been performed in order, first, to reduce their contributions in
each event, and then to subtract the remaining contributions event-by-event 
in the process of jet finding. For the former, HIJING simulations have shown 
that the application of a p$_T$ cut of 1 or 2~GeV/c on charged particles and the 
reduction of the jet cone size to R~=~0.3-0.4 have to be applied. For the later, 
three methods of subtraction (statistic, cone and ratio) 
have been tested and compared~\cite{estienne3,estienne6}.
With 
the iterative jet cone finder based on the UA1 cone algorithm and a full detector simulation, 
a mean cone energy of $\mean{E}$~=~45 GeV and a resolution of $RMS/\mean{E}\sim $40\% 
are obtained using only charged particles in the reconstruction of 100~GeV mono-energetic 
jets in $p$-$p$ collisions. The inclusion of neutral particles measured in the ALICE calorimeter~\cite{estienne7} 
allows one to increase $\mean{E}$ to 70~GeV while improving the resolution up to $\sim$ 30\% 
(for a jet cone radius of R~=~0.4). A deeper study of background subtraction in
Pb-Pb collisions has shown that the jet reconstruction algorithms bias the mean background energy 
contained in the jet cone toward values higher than the ones found in the same area outside the
jet. This bias leads to an over estimation of the reconstructed
energy~\cite{estienne8}. We obtain a preliminary $\mean{E}\approx$~75~GeV and a
resolution of 30-35\% for the case charged + neutral particles for 100~GeV 
mono-energetic jets in minimum-bias Pb-Pb collisions.\\

The full jet spectrum has also been simulated and studied in $p$-$p$ and Pb-Pb
collisions. Due to the steeply falling jet production spectrum and the
background and signal fluctuations of the reconstructed jet energy, we
observe a smearing of the reconstructed spectra in $p$-$p$. This effect is
increased in Pb-Pb due to the background fluctuations and its bias in
the jet finding process. Jets clearly present advantages for the study of 
jet quenching at the LHC compared to the studies of leading particles 
which are trigger- and surface-biased. With its reconstruction
capabilities, ALICE can reconstruct jets with enough
resolution for Pb-Pb studies even though the observed smearing of the
reconstructed spectrum, due to the steeply falling shape, 
will have to be well controlled both in elementary and Pb-Pb collisions.


\tit{Parton fragmentation studies in ATLAS}

\auth{Ji\v{r}\'i Dolej\v{s}\'i, for ATLAS collaboration}

ATLAS installation and commissioning proceeds towards ability to record first $pp$ 
collisions during the year 2008.  The ATLAS heavy ion working group has examined 
the detector performance in conditions of Pb-Pb collision at 5.5 TeV per nucleon, 
based on full simulations of the detector response, and continues in detailed 
studies~\cite{dolejsi1,dolejsi2}. The focus of this talk is on the jet studies.
A dedicated jet finder package has been developed and integrated into the standard 
ATLAS simulation framework. It contains the background subtraction, where the 
background is estimated over the region outside jet candidate, and iteratively uses 
the jet cone algorithm on the calibrated energy depositions in individual cells of the 
ATLAS calorimeters. This package has been tuned on PYTHIA jets embedded into HIJING 
Pb-Pb collision with a hard-cut of high $p_T$ processes. 
The simulations show a reconstruction efficiency of  the cone algorithm steeply 
rising from about 50\%  at $E_T$ = 60~GeV  up to 90\% above 90~GeV. 
The fake fraction is about 0.04 at $E_T$ = 60~GeV and decreases with higher $E_T$. 
These results were obtained for events with moderately high d$N$/d$\eta$ = 2700. 
The axis of the jet can be measured with an angular resolution of about 0.05 (both 
in $\phi$ and $\eta$) for jets with $E_T$ = 60~GeV and improves for higher $E_T$'s. 
The energy resolution is about 30\% at 60~GeV and is flat over the accessible $\eta$ range.\\

The precise position resolution enables a reliable extraction of the jet shapes. The distributions of 
$j_T$ (particle momentum perpendicular to the jet axis) as well as jet fragmentation functions 
$D(z)$ have also been studied using the tracks reconstructed in the ATLAS Inner Detector. 
The tracking performance was found to be sufficiently good for these distributions. The 
agreement between the true and reconstructed distributions is good, providing that the 
tracks from the underlying event are subtracted. The fast-$K_T$ algorithm~\cite{estienne5} is 
studied as one of the possible alternatives. Its efficiency and energy resolution is comparable 
to that of the cone algorithm and more detailed studies continue.

\tit{Extracting fragmentation functions with $\gamma$-tagged jet events in Pb-Pb  with the CMS experiment}

\auth{Christof Roland, for the CMS Collaboration}

\newcommand {\et}          {\ensuremath{E_{\mathrm{T}}}}
\newcommand {\PbPb}        {Pb-Pb}
\newcommand {\abs}[1]      {\ensuremath{\left| #1 \right|}}

The energy loss of fast partons traversing the strongly interacting matter
produced in high-energy nuclear collisions is one of the most interesting
observables to probe the nature of the produced medium. The collisional and
radiative energy loss of the partons will modify their fragmentation functions
depending on the path length of the partons in the medium and the medium density.
Pb-Pb collisions at $\sqrt{s_{\scriptscriptstyle{{\rm NN}}}}=5.5$~TeV at the LHC
will open access to the first detailed measurements of the in-medium modifications
of fragmentation functions of parton initiated jets.
The fragmentation functions can be studied using the $\gamma$-jet
channel~\cite{Wang:1996yh,Arleo:2006xb}, where the  initial transverse energy of the
fragmenting parton can be determined from the photon transverse energy, since the
photon does not strongly interact with the medium.\\

The CMS experiment features large acceptance electromagnetic and hadronic calorimeters,
and a high precision tracking system well suited to nucleus-nucleus collisions at the LHC~\cite{D'Enterria:2007xr}.
To extract fragmentation functions of parton initiated jets in $\gamma$-jet events,
isolated photons reconstructed in the electromagnetic calorimeter ($\abs{\eta}<2.0$) are selected and
correlated with back-to-back reconstructed calorimeter jets ($\abs{\eta}<2.0$) in \PbPb\ events.
For the selected photon jet pairs, charged particles reconstructed in the tracking system ($\abs{\eta}<2.5$)
that lie within a $0.5$ radius cone in $\eta$-$\phi$ around the reconstructed jet axis
are selected. The fragmentation function is constructed correlating the transverse
energy of the photons, with the transverse momentum of the tracks in the jet cone.\\

This study based on a full GEANT simulation of a data set of $\gamma$-jet events,
containing a $\gamma$ with \et~$> 70$~GeV, embedded in central \PbPb\ events
corresponding to a full year of data taking with an integrated luminosity of 0.5~nb$^{-1}$.
The capability to measure jet fragmentation functions in the $\gamma$-jet channel with high
statistics is demonstrated. The systematic uncertainties are estimated to be on the order of
30\% (10\%) for quenched events containing a photon with \et\ $> 70$~GeV ($100$~GeV) mostly
due to a bias induced by the finite efficiency of the jet reconstruction. Overall the
fragmentation function can be reconstructed with high significance for 0.2 $< \xi <$ 5
for both, \et\ $> 70$~GeV and $> 100$~GeV.
Reconstructed $\gamma$-jet events can be used to study the dependence of
high-$\pt$ parton fragmentation on the medium and will provide
sensitivity to the foreseeable changes in the fragmentation functions
relative to vacuum fragmentation. This measurement will allow a
quantitative test of proposed mechanisms for medium-induced parton energy loss, 
testing fundamental properties of high-density QCD. 

\tit{Issues of fragmentation function and medium effects in single inclusive 
production and two particle correlations in $p$-$p$ and A-A collisions}

\auth{Michael Tannenbaum}

Hard-scattering of point-like constituents (partons) in $p$-$p$ collisions was discovered at the 
CERN-ISR in 1972 in measurements of inclusive single or pairs of hadrons with large 
transverse momentum~($p_T$). Due to the steeply falling power-law ${p}_T$  spectrum of the 
hard-scattered partons, the inclusive single particle (e.g. $\pi^0$) $p_{T_t}$ spectrum from jet 
fragmentation of a parton with $\hat{p}_{T_t}$ is dominated by trigger fragments with large $\mean{z_t}\sim 0.7-0.8$,
where $z_t=p_{T_t}/\hat{p}_{T_t}$ is the fragmentation variable. It was generally assumed, 
following Feynman, Field and Fox~\cite{tannenbaum1}, as shown by data from the ISR experiments, that 
the $p_{T_a}$ distribution of away side hadrons from a single particle trigger [with $p_{T_t}$], 
corrected for $\mean{z_t}$, would be the same as that from a jet-trigger and follow the same 
fragmentation function measured in $e^+ e^-$ or DIS.
PHENIX~\cite{tannenbaum2,tannenbaum3} attempted to measure the fragmentation function from the away side 
$x_E \sim p_{T_a}/p_{T_t}$ distribution of charged particles triggered by a $\pi^0$ in $p$-$p$ collisions 
and showed by explicit calculation that the $x_E$ distribution is actually quite insensitive to the 
fragmentation function.  A new formula for the distribution of an associated
away-side particle with transverse momentum $p_{T_a}$, which is
presumed to be a fragment of an away-jet with $\hat{p}_{T_a}$, with exponential fragmentation 
function $D(z)=Be^{-bz}$, triggered by a particle with transverse momentum $p_{T_t}$,
presumably from a trigger-side jet with $\hat{p}_{T_t}$, invariant cross section, $Ed^3\sigma/dp^3=A\hat{p}_{T_t}^{-n}$, was
given~\cite{tannenbaum2}: \  ${dP_{p_{T_a}}/dx_E}|_{p_{T_t}}\approx
{{\langle{m}\rangle}\over\hat{x}_h} {(n-1)\over
{(1+ {x_E /{\hat{x}_h}})^{n}}}$.  
This formula relates $x_E$, the ratio of the transverse
momenta of the measured particles, to $\hat{x}_h=\hat{p}_{T_a}/\hat{p}_{T_t}$,
the ratio of the transverse momenta of the away-side to
trigger-side jets, where $\langle m\rangle$ is the mean
multiplicity of particles in the away jet.\\

Many analyses of the
away-jet $p_{T_a}$ distributions in Au-Au collisions are
available; but these tend to describe the effect of the medium
with the variable $I_{AA}(x_E)$, the ratio of the $x_E$
distribution in A-A collisions to that in $p$-$p$ collisions, which
typically shows an enhancement at low values of $x_E$ and a
suppression at higher values of $x_E$. Such behavior could be
explained as a decrease in $\hat{x}_h$ in A-A collisions due to
energy loss of the away jet in the medium. Fits of the above
formula to the available data were presented to establish
whether: a) the away-jets simply lose energy; b) some of the
away-jets lose energy, others punch-through without losing
energy; etc.\\

Measurements of direct single $\gamma$ production in $p$-$p$ and Au-Au are also 
discussed. Fragmentation photons in $p$-$p$ collisions, if they exist, can be
eliminated by isolation cuts, although, at the ISR and at RHIC, measurements of correlations to a 
direct single $\gamma$ show very little activity, if any, on the same side. In contrast to triggers 
with $\pi^0$ which are the fragments of jets, away-side correlations for direct single $\gamma$'s 
which are direct participants in 2-2 hard scattering should measure the fragmentation function of 
the away side jet, modulo $k_T$ smearing. This fragmentation function can be plotted conventionally 
as a function of $x_E$ or as a function of $\xi=\ln(1/x_E)$ as suggested for jets at the LHC~\cite{Borghini:2005em}.

\tit{Fragmentation function from direct-photon associated yields at RHIC}

\auth{Jan Rak}

\def\kt#1{${k_{\rm T#1}}$}
\def\sqrtrms#1{$\sqrt{\la #1^{2} \ra}$}
\def\la{\left< }
\def\ra{\right> }
\newcommand{\mzt} {$\la z_{\rm t} \ra$}

The interest of heavy-ion community in the fragmentation function (FF) is manifold.
The detailed knowledge of the FF modification in heavy-ion collisions
is important for constraints of various jet-quenching models~\cite{Borghini:2005em}. Another region of interest
is related to the \pt{}-broadening predicted to accompany the parton interaction with excited nuclear 
medium~\cite{Ivan}. One way of evaluating the \pt{}-broadening is to measure the parton intrinsic 
momentum \kt{}~\cite{tannenbaum2}. It has been shown that in order to extract the value of \sqrtrms{k_{\rm T}}
the knowledge of the average trigger particle momentum fraction \mzt\ and thus FF is vital 
(see Eq. (22) in~\cite{tannenbaum2}). It has been also demonstrated that, despite a common belief, 
an analysis of charged particle distributions associated with the high-\pt{} trigger hadron cannot provide
sufficient constraint for the experimental determination of FF. The reason for that luck of sensitivity 
of high-\pt{}\ trigger hadron associated spectra to the shape of FF lies in the fact that when events 
with different associated transfer momenta are selected the trigger particle momentum fraction 
\mzt\ changes event though the trigger particle momentum is fixed.  
In this talk I have presented an analysis of direct-photon associated spectra which should be free 
from the above mention problem. There is, however, an effect of \kt{}-smearing which causes 
the imbalance between the photon-quark momenta. I have presented a method of unfolding 
the \kt{}-smearing which allows to recover the shape of FF from photon associated spectra.

\tit{Systematics of complete fragment distributions from nuclear collisions}

\auth{Thomas A. Trainor}

The study of fragmentation in RHIC nuclear collisions addresses the central question how formation 
and evolution of a QCD medium in heavy ion collisions may affect parton scattering and fragmentation. 
Whereas previous emphasis has been placed on larger parton energy scales described by perturbative methods, 
we seek a complete description of all fragmentation processes in nuclear collisions.
Fragmentation in elementary collisions provides a reference system. 
To better determine the energy scale dependence of fragmentation functions from $e^+e^-$,
the FFs were replotted on a normalized rapidity variable, resulting in a 
compact form precisely represented by the beta distribution, its two parameters varying slowly and 
simply with parton energy scale $Q$~\cite{lepmini}. The new parameterization extrapolates 
fragmentation functions to small $Q$ and describes the fragmentation process down to small 
transverse momentum $p_T$ in nuclear collisions at RHIC.
Analysis of transverse momentum $p_T$ spectra to $p_T = 6$~GeV/c for 10 multiplicity 
classes of $p$-$p$ collisions at $\sqrt{s} = 200$~GeV revealed that the spectrum shape can be decomposed
into a part with amplitude proportional to multiplicity and described by a L\'evy distribution on transverse 
mass $m_T$, and a part with amplitude proportional to multiplicity squared and described by a Gaussian 
distribution on transverse rapidity $y_T$~\cite{ppprd}. The functional forms of the two parts are nearly 
independent of event multiplicity, and they can be identified with the soft and hard components of a 
two-component model of $p$-$p$ collisions: longitudinal participant nucleon fragmentation and transverse scattered 
parton fragmentation. Interpretation of the hard component as parton fragmentation is supported by 
correlation studies~\cite{tomismd,minijets}.\\

A similar two-component analysis of spectra to $p_T = 12$~GeV/c for identified pions and protons from 
200~GeV Au-Au collisions applied to Au-Au centrality dependence revealed that the soft-component 
reference is again a L\'evy distribution on transverse mass $m_T$, and the hard-component reference 
is again a Gaussian on $y_T$, but with added exponential ($p_T$ power-law) tail~\cite{2comp}. 
Deviations of data from the two-component reference are described by hard-component ratio $r_{AA}$ 
which generalizes nuclear modification factor $R_{AA}$. Centrality evolution of pion and proton spectra 
is dominated by changes in parton fragmentation. The structure of $r_{AA}$ suggests that parton energy 
loss produces a negative boost $\Delta y_T$ of a large fraction (but not all) of the minimum-bias fragment 
distribution (hard component), and that lower-energy partons suffer relatively less energy loss, possibly 
due to color screening.\\

A recent analysis of azimuth correlations based on methods in~\cite{flowmeth,gluequad} has isolated 
azimuth quadrupole components from published $v_2(p_T)$ data (called elliptic flow) as spectra on 
transverse rapidity $y_T$ for identified pions, kaons and lambdas from minimum-bias Au-Au collisions 
at 200~GeV. The form of the spectra reveals that the three hadron species are emitted from a common 
source with boost $\Delta y_{t0} \sim 0.6$. The quadrupole spectra have a L\'evy form on $m_T$ 
similar to the soft components of single-particle spectra, but with significantly reduced ($\sim 0.7\times$) 
slope parameter $T$. Comparison of quadrupole spectra with single-particle spectra suggests that the 
quadrupole component comprises a small fraction ($< 5$\%) of the total hadron yield, contradicting 
the hydrodynamic picture of a thermalized, flowing bulk medium. The form of $v_2(p_T)$ is, within a 
constant factor, the product of $p'_t$ ($p_T$ in the boost frame) times the ratio of the quadrupole 
spectrum to the single-particle spectrum. The form of the spectrum ratio then implies that above 
0.5~GeV/c, $v_2(p_T)$ is dominated by the hard component of the single-particle spectrum (minijets). 
It is thus unlikely that so-called {\em constituent-quark scaling} attributed to $v_2$ is 
relevant to soft hadron production mechanisms (e.g., chemical freezeout). These results suggest 
instead that ``elliptic flow'' may result from fragmentation of gluonic quadrupole radiation, not 
hydrodynamic evolution.

\appendix

\newpage

\section{List of participants}

\begin{wparticipants}
\parbox[t]{0.55\textwidth}{
\spk{S.~Albino}{Universit\"at Hamburg},\\
\spk{F.~Anulli}{INFN, Roma},\\
\spk{E.~Aschenauer}{DESY, Zeuthen},\\
\spk{F.~Arleo}{LAPTH, Annecy},\\
\spk{D.~Besson}{Kansas University},\\
\spk{N.~Borghini}{Heidelberg Univ.},\\
\spk{W.~Brooks}{Santa Maria Univ., Valparaiso},\\
\spk{B.~Buschbeck}{OA Wien, Austria},\\
\spk{M.~Cacciari}{LPTHE, Universit\'e Paris 6,},\\
\spk{E.~Christova}{Inst. Nuc. Res. \& Nuc Energy, Sofia},\\
\spk{G.~Corcella}{Centro Fermi Roma, SNS/INFN Pisa},\\ 
\spk{D.~d'Enterria}{CERN, Geneva},\\
\spk{J.~Dolej\v{s}\'i}{Charles University, Prague,},\\
\spk{S.~Domdey}{Universit\"at Heidelberg},\\
\spk{M.~Estienne}{IPHC, Strasbourg},\\
\spk{O.~Fochler}{University Frankfurt},\\
\spk{T.~Gousset}{Subatech, Nantes},\\
\spk{K.~Hamacher}{Bergische Univ. Wuppertal},\\
\spk{M.~Heinz}{Yale University},\\
\spk{K.~Hicks}{Ohio University},\\
\spk{D.~Kettler}{Univ. of Washington, Seattle},\\
\spk{S.~Kumano}{KEK, Japan},\\
\spk{B.~Machet}{	LPTHE, Univ. Paris 6},\\
\spk{G.~Milhano}{CENTRA-IST, Lisbon},\\
}
\parbox[t]{0.45\textwidth}{
\spk{S.-O.~Moch}{DESY, Zeuthen},\\
\spk{V.~Muccifora}{INFN-LNF, Frascati},\\
\spk{S.~Pacetti}{Centro Fermi Roma, LNF Frascati},\\
\spk{R.~P\'erez-Ramos}{Universit\"at Hamburg},\\
\spk{H.-J.~Pirner}{Univ. Heidelberg},\\
\spk{A.~Pronko}{Fermilab, Chicago},\\
\spk{J.~Putschke}{Yale University},\\
\spk{M.~Radici}{INFN - Sezione di Pavia},\\
\spk{J.~Rak}{University of Jyv\"askyl\"a,},\\
\spk{C.~Roland}{MIT, Cambridge, MA},\\
\spk{G.~Rudolph}{Universit\"at Innsbruck},\\
\spk{Z.~R\'urikov\'a}{DESY, Hamburg },\\
\spk{C.A.~Salgado}{Univ. Santiago de Compostela},\\
\spk{S.~Sapeta}{Jagellonian U. Cracow},\\
\spk{D.H.~Saxon}{University of Glasgow},\\
\spk{R.~Seidl}{RIKEN BNL, Upton, NY},\\
\spk{R.~Seuster}{Victoria Univ., BC},\\
\spk{M.~Stratmann}{RIKEN, Japan},\\
\spk{M.J.~Tannenbaum}{BNL, Upton, NY},\\
\spk{M.~Tasevsky}{Charles Univ., Prague},\\
\spk{T.~Trainor}{University of Washington, Seattle},\\
\spk{D.~Traynor}{QMUL, London},\\
\spk{M.~Werlen}{	LAPTH, Annecy},\\
\spk{C.~Zhou}{McGill Univ., Montreal}
}
\end{wparticipants}

\section{Programme}

\begin{tabbing}
xx:xx \= A very very very long name \= \kill
{\bf Monday, 25 February 2008}\\\vspace{0.7cm}
09:00 \hspace{0.4cm} \> {\it Welcome and introduction} (10')  \> \hspace{7.65cm}  F. Arleo, D. d'Enterria \\
09:10 \hspace{0.4cm} \> {\it Sources of uncertainty in quark \& gluon FFs into hadrons \& photons} (50')   \> \hspace{7.65cm} M. Werlen \\
10:00 \hspace{0.4cm} \> {\it OPAL results on quark, gluon fragmentation} (30')   \> \hspace{7.65cm} M. Tasevsky \\
11:00 \hspace{0.4cm} \> {\it Colour flux studies in quark \& gluon Fragmentation in DELPHI} (30')   \> \hspace{7.65cm} B. Buschbeck \\
11:30 \hspace{0.4cm} \> {\it ALEPH results on quark and gluon fragmentation} (30')   \> \hspace{7.65cm}  G. Rudolph\\
12:00 \hspace{0.4cm} \> {\it Colour coherence and a comparison of gluon and quark fragmentation} (30')   \> \hspace{7.65cm}  K. Hamacher \\
14:30 \hspace{0.4cm} \> {\it AKK fragmentation functions: latest developments.} (40')    \> \hspace{7.65cm} S. Albino\\
15:10 \hspace{0.4cm} \> {\it DSS Fragmentation functions} (40')   \> \hspace{7.65cm} M. Stratmann\\
16:20 \hspace{0.4cm} \> {\it HKNS fragmentation functions and proposal for exotic-hadron search} (40')   \> \hspace{7.65cm}  S. Kumano \\
17:00 \hspace{0.4cm} \> {\it Low-$Q^2$ fragmentation functions in $e^+e^-$ collisions} (40')   \> \hspace{7.65cm} D.  Kettler\\\\\\\\\\

{\bf  Tuesday 26 February 2008 }\\
08:30 \hspace{0.4cm} \> {\it Evidence for power corrections in heavy quark fragmentation in $e^+e^-$} (30')  \> \hspace{7.65cm} M.~Cacciari\\
09:00 \hspace{0.4cm} \> {\it Heavy Quark Fragmentation at 10.6 GeV} (30')  \> \hspace{7.65cm}  R.~Seuster\\
09:30 \hspace{0.4cm} \> {\it Initial state radiation at BaBar} (30')   \> \hspace{7.65cm} S.~Pacetti \\
10:00 \hspace{0.4cm} \> {\it Heavy quark fragmentation with an effective coupling constant} (30')   \> \hspace{7.65cm}  G.~Corcella\\
11:00 \hspace{0.4cm} \> {\it Charm fragmentation at HERA} (30')   \> \hspace{7.65cm} Z.~R\a'urikov\a'a\\
11:30 \hspace{0.4cm} \> {\it Inclusive light and charmed particle production in BaBar} (30')   \> \hspace{7.65cm}  S.~Pacetti \\
14:00 \hspace{0.4cm} \> {\it Discussion: interfacing fragmentation functions ?} (30')   \> \hspace{7.65cm}  F. Arleo, D. d'Enterria \\
14:30 \hspace{0.4cm} \> {\it Experimental constraints on strange quark fragmentation functions} (30')    \> \hspace{7.65cm} M.~Heinz \\
15:00 \hspace{0.4cm} \> {\it Latest developments on parton to kaon fragmentation} (30')   \> \hspace{7.65cm} E.~Christova\\
15:30 \hspace{0.4cm} \> {\it Inclusive $K^0_sK^0_s$  resonance production in ep collisions at HERA} (30')   \> \hspace{7.65cm}  Changyi Zhou\\
16:30 \hspace{0.4cm} \> {\it Theory of (extended) dihadron fragmentation functions } (30')    \> \hspace{7.65cm} M.~Radici\\
17:00 \hspace{0.4cm} \> {\it Time-like splitting functions at NNLO in QCD} (30')   \> \hspace{7.65cm}  Sven-Olaf Moch  \\\\

{\bf Wednesday 27 February 2008 }\\
09:00 \hspace{0.4cm} \> {\it Recent developments in parton fragmentation at MLLA and beyond} (45')  \> \hspace{7.65cm} R.~P\a'erez Ramos \\
09:45 \hspace{0.4cm} \> {\it Studies of Jet Fragmentation at CDF} (30')   \> \hspace{7.65cm} A.~Pronko\\
10:45 \hspace{0.4cm} \> {\it Particle production at ZEUS } (40')   \> \hspace{7.65cm} D.H. Saxon\\
11:25 \hspace{0.4cm} \> {\it Particle production and fragmentation studies in H1} (30')   \> \hspace{7.65cm} D. Traynor\\
11:55 \hspace{0.4cm} \> {\it Review on parton fragmentation studies in BaBar} (30')   \> \hspace{7.65cm} F.~Anulli \\
14:30 \hspace{0.4cm} \> {\it Fragmentation Studies at CLEO} (30')   \> \hspace{7.65cm} D. Besson\\
15:00 \hspace{0.4cm} \> {\it Light quark and spin dependent fragmentation at Belle} (30')    \> \hspace{7.65cm} R.~Seidl\\
16:00 \hspace{0.4cm} \> {\it HERMES: News on fragmentation from nucleons to nuclei } (30')    \> \hspace{7.69cm}E.~Aschenauer \\
16:30 \hspace{0.4cm} \> {\it $p_T$ broadening in nuclear collisions} (30')   \> \hspace{7.65cm} H.-J.~Pirner\\\\

{\bf Thursday 28 February 2008 }\\
09:00 \hspace{0.4cm} \> {\it Medium modified Fragmentation Functions: an overview} (45')  \> \hspace{7.65cm} C.~Salgado\\
09:45 \hspace{0.4cm} \> {\it Parton fragmentation studies in ATLAS } (30')   \> \hspace{7.65cm} J.~Dolej\v{s}\a'i \\
10:45 \hspace{0.4cm} \> {\it Fragmentation functions with $\gamma$-jet in Pb-Pb at 5.5 TeV at CMS} (30')   \> \hspace{7.65cm} C.~Roland\\
11:15 \hspace{0.4cm} \> {\it Reconstructing jets in Pb-Pb collisions in ALICE } (30')   \> \hspace{7.65cm}  M.~Estienne\\
11:45 \hspace{0.4cm} \> {\it Jet fragmentation studies in the ALICE experiment} (30')   \> \hspace{7.65cm}   J.~Putschke\\
14:30 \hspace{0.4cm} \> {\it Systematics of complete fragment distributions in nuclear collisions} (30')   \> \hspace{7.65cm}  T.~Trainor\\
15:00 \hspace{0.4cm} \> {\it Hadronic composition as a characteristics of jet quenching at LHC } (30')   \> \hspace{7.65cm}  S.~Sapeta\\
16:00 \hspace{0.4cm} \> {\it Fragmentation function from direct-photon associated yields at RHIC } (30')    \> \hspace{7.65cm} J.~Rak\\
16:30 \hspace{0.4cm} \> {\it FFs and medium effects in single \& 2-particle production in $p$-$p$ and A-A} (30')   \> \hspace{7.75cm} M.J. Tannenbaum\\\\

{\bf Friday 29 February 2008 }\\
09:30 \hspace{0.4cm} \> {\it Attenuation of leading hadrons in DIS on nuclei: an overview } (45')  \> \hspace{7.65cm} V.~Muccifora \\
10:15 \hspace{0.4cm} \> {\it Hadron Formation in Semi-Inclusive DIS on Nuclei at HERMES } (30')   \> \hspace{7.65cm}  G.~Elbakyan\\
11:15 \hspace{0.4cm} \> {\it $\pi,K$ hadronization from Electroproduction DIS on Nuclei at CLAS} (30')   \> \hspace{7.65cm} K. Hicks\\
11:45 \hspace{0.4cm} \> {\it Parton propagation \& hadron formation: current status, future prospects} (30')   \> \hspace{7.75cm} W. Brooks\\
12:15 \hspace{0.4cm} \> {\it QCD evolution of jets in the quark-gluon plasma } (30')   \> \hspace{7.65cm} S. Domdey\\

\end{tabbing}

%


\newpage


\end{document}